\documentclass{jpp}

\usepackage{graphicx}
\usepackage{natbib}
\usepackage{amsmath}
\usepackage{amssymb}
\usepackage{mathtools}
\usepackage{lscape}
\usepackage{color}  
\usepackage{rotating}
\usepackage{booktabs}
\usepackage{caption}
\usepackage{ulem}
\usepackage{psfrag}
\usepackage{pstool}
\usepackage[colorlinks,urlcolor=blue,citecolor=blue,linkcolor=blue]{hyperref}
\usepackage{ulem}

\ifCUPmtlplainloaded \else
  \checkfont{eurm10}
  \iffontfound
    \IfFileExists{upmath.sty}
      {\typeout{^^JFound AMS Euler Roman fonts on the system,
                   using the 'upmath' package.^^J}%
       \usepackage{upmath}}
      {\typeout{^^JFound AMS Euler Roman fonts on the system, but you
                   dont seem to have the}%
       \typeout{'upmath' package installed. JPP.cls can take advantage
                 of these fonts, if you use 'upmath' package.^^J}%
      }
  \else
  \fi
\fi

\ifCUPmtlplainloaded \else
  \checkfont{msam10}
  \iffontfound
    \IfFileExists{amssymb.sty}
      {\typeout{^^JFound AMS Symbol fonts on the system, using the
                'amssymb' package.^^J}%
       \usepackage{amssymb}%
         \let\leq=\leqslant
       \let\ge=\geqslant  \let\geq=\geqslant
      }{}
  \fi
\fi

\ifCUPmtlplainloaded \else
  \IfFileExists{amsbsy.sty}
    {\typeout{^^JFound the 'amsbsy' package on the system, using it.^^J}%
     \usepackage{amsbsy}}
    {}
\fi

\newsavebox{\astrutbox}
\sbox{\astrutbox}{\rule[-5pt]{0pt}{20pt}}

\def\<{\langle}
\def\>{\rangle}

\def\bi{\bf}

\def\be{\begin{equation}}
\def\ee{\end{equation}}
\def\bea{\begin{eqnarray}}
\def\eea{\end{eqnarray}}

\newcommand{\ms}{\noalign{\vspace{3pt plus2pt minus1pt}}}

\newfont{\myfont}{cmmib10}

\title[Wave dispersion in pulsar plasma 2]{Wave dispersion in pulsar plasma:\\ 2. Pulsar frame}

\author[M. Z. Rafat, D. B. Melrose and A. Mastrano]%
{M.\ns Z. \ns R\ls A\ls F\ls A\ls T$^{1}$ \ns
D.\ns B.\ns M\ls E\ls L\ls R\ls O\ls S\ls E$^1$%
  \thanks{Email address for correspondence: donald.melrose@sydney.edu.au},\ns \and
A.\ns  M\ls A\ls S\ls \ls T\ls R\ls A\ls N\ls O$^{1}$}
\affiliation{$^1$SIfA, School of Physics, The University of Sydney, NSW 2006, Australia
}

\pubyear{2017}
\volume{?}
\pagerange{?}
\date{?; revised ?; accepted ?. - To be entered by editorial office}
\begin{document}

\maketitle

\begin{abstract}
Wave dispersion in a pulsar plasma is discussed emphasizing the relevance of different inertial frames, notably the plasma rest frame ${\cal K}$ and the pulsar frame ${\cal K}'$ in which the plasma is streaming with speed $\beta_{\rm s}$. The effect of a Lorentz transformation on both subluminal, $|z|<1$, and superluminal, $|z|>1$, waves is discussed. It is argued that the preferred choice for a relativistically streaming distribution should be a Lorentz-transformed J\"uttner distribution; such a distribution is compared with other choices including a relativistically streaming Gaussian distribution. A Lorentz transformation of the dielectric tensor is written down, and used to derive an explicit relation between the relativistic plasma dispersion functions in ${\cal K}$ and ${\cal K}'$. It is shown that the dispersion equation can be written in an invariant form, implying a one-to-one correspondence between wave modes in any two inertial frames. Although there are only three modes in the plasma rest frame, it is possible for backward-propagating or negative-frequency solutions in ${\cal K}$ to transform into additional forward-propagating, positive-frequency solutions in ${\cal K}'$ that may be regarded as additional modes.
\end{abstract}

\begin{PACS}
\end{PACS}

\section{Introduction}

In an accompanying paper (Rafat et al 2019, hereinafter Paper~1) we discuss wave dispersion in the rest frame, denoted ${\cal K}$, of a pulsar plasma emphasizing the importance of the intrinsic spread in electron (and positron) energies, $\langle\gamma\rangle > 1$, with $\langle\gamma\rangle \gg 1$ on a highly relativistic plasma. In this paper we discuss aspects of the plasma physics that involve Lorentz transforming between frames.  In particular, we consider the effects of the Lorentz transformation between ${\cal K}$ and the pulsar frame, ${\cal K}'$, in which the plasma is streaming outwards at speed $\beta_{\rm s}$, where we use ``speed'' to refer to a velocity component along the direction of the magnetic field relative to the speed of light. Our ultimate objective (in Paper~3) is to discuss possible wave growth leading to pulsar radio emission. Identification of the wave properties in ${\cal K}$ (Paper~1) is one important ingredient needed for such a discussion. However, the relevant frame when treating the wave growth, the emission process and the escape of radiation is ${\cal K}'$. The transformation of the wave properties between the two frames includes the transformation of the relativistic plasma dispersion function (RPDF) $W(z)$ in ${\cal K}$ to $W'(z')$ in ${\cal K}'$. As in Paper~1, we describe the wave dispersion in terms of the frequency $\omega$, phase speed $z=\omega/k_\parallel c$ and angle $\theta$ of propagation. The Lorentz transformation relates $\omega$, $z$ and $\theta$ in ${\cal K}$ to $\omega'$, $z'$ and $\theta'$ in ${\cal K}'$. Another important aspect concerns the choice of distribution function. As in Paper~1 we suggest that the default choice for a relativistic distribution of particles in ${\cal K}$ should be a 1D J\"uttner distribution. Here we argue that the default choice for the distribution function for a beam is that obtained by applying a Lorentz transform to a 1D J\"uttner distribution in ${\cal K}$. Alternative choices for a relativistic distribution function in ${\cal K}$ include a power-law \citep[][\S17]{KT73}, a relativistic Gaussian \citep{LP82,AM98} and water-bag \citep{AB86} and bell \citep{GMG98} distributions. For non-streaming distributions the effects of the different choices is primarily on the form of the RPDF, and these effects are relatively minor \citep{GMG98}. However, different choices have a much larger effect for streaming distributions. We find that the Lorentz-transformed distribution function is very much broader than the streaming Gaussian distribution usually assumed. This has major implications for beam-driven instabilities discussed in Paper~3. One example that we discuss in detail concerns the ``separation'' condition, that is, the condition for two relatively streaming distributions to be separated (in 4-speed $u=\gamma\beta$), rather than overlapping, so that one can be identified as a beam propagating through the other (the background).\footnote{We remark that a distribution of highly relativistic particles may also be present, but we neglect this here because the growth rate due to such particles is known to be too small to be relevant. Similarly we neglect any nonthermal high-energy tail on the background (Lorentz-transformed J\"uttner) distribution on the grounds that it cannot dominate in determining the growth rate, unlike the nonrelativistic case, for example, modeled as a kappa-distribution in treating solar radio bursts \citep[e.g.,][]{CAIRNS2017549}.}

Wave dispersion in ${\cal K}'$ may be treated using three different (but equivalent) approaches. One approach is to treat the wave dispersion in ${\cal K}$ and Lorentz transform the wave solutions to ${\cal K}'$. Two effects of the Lorentz transformation on a wave are well-known in the context of escaping pulsar radio emission: the effect (Lorentz boost) on the  frequency \citep{Letal98} and the effect (aberration) on the direction of propagation \citep{Cordes78,GG03}. The transformation of the phase speed is a trivial application of the relativistic addition of velocities, $z'=(z+\beta_{\rm s})/(1+z\beta_{\rm s})$, but some care is needed in the application to wave dispersion because either $\omega'$ or $z'$ may be opposite in sign to $\omega$ or $z$. Formally, $\omega'<0$ may be treated by using the symmetry of the dispersion equation under $\omega',{\bi k}'\to-\omega',-{\bi k}'$ to relate the positive- and negative-frequency solutions, by requiring that the physical solution of the dispersion relation (in any frame) correspond to a positive frequency in that frame. Other approaches involve deriving the wave dispersion directly in ${\cal K}'$, with two alternatives relating to the way the dielectric tensor is identified in ${\cal K}'$. One way is to Lorentz transform the distribution function and use the transformed distribution function in calculating the dielectric tensor in ${\cal K}'$. The other way is to Lorentz transform the dielectric tensor from ${\cal K}$ to ${\cal K}'$. The latter approach involves transforming the relativistic plasma dispersion function (RPDF) $z^2W(z)$ in ${\cal K}$ to $z'^2W'(z')$ in ${\cal K}'$. We establish the equivalence of these approaches in general, by showing that the dispersion equation may be written in invariant form. We also illustrate the equivalence for specific wave modes.

The equivalence of the two ways of relating wave dispersion in ${\cal K}$ and ${\cal K}'$ implies there are the same number (three modes) in both ${\cal K}$ (Paper~1) and ${\cal K}'$. However, in principle (up to three) additional modes can arise in ${\cal K}'$ from a re-interpretation of backward-propagating or negative-frequency solutions in ${\cal K}$ transforming into a forward-propagating, positive-frequency solution in ${\cal K}'$. Other authors identified a fourth mode \citep{BGI93,Isotmin01,LG-S06} in ${\cal K}'$, and we argue that this is partly the result of such an interpretation. However, this point is confused by what we suggest is an inappropriate approximation made effectively in evaluating the RPDF in ${\cal K}'$.

In \S\ref{sect:Lorentz} we write down the Lorentz transformation between ${\cal K}$ and ${\cal K}'$ for a wave and also for a 1D distribution function. In \S\ref{sect:beam} we argue that a beam should be modeled as a Lorentz-transformed J\"uttner distribution, and we introduce a multi-beam model composed of several such distributions. In \S\ref{sect:separation} we estimate the separation condition for two such (relatively streaming) distributions to be regarded as non-overlapping, and point out that this condition is more restrictive than might be anticipated. We write down the Lorentz transformation of the dielectric tensor and of the dispersion equation in \S\ref{app:LT} and derive an explicit relation between the RPDFs in ${\cal K}$ and in ${\cal K}'$. We discuss the transformed dispersion equation and dispersion relations in \S\ref{sect:dielectric}. In \S\ref{sect:fourth} we show how each of the three modes in ${\cal K}$ splits into two branches in ${\cal K}'$ and we comment on the suggested fourth mode in ${\cal K}'$. We discuss our results and summarize our conclusions in \S\ref{sect:conclusions}.

\section{Lorentz transformation between rest and pulsar frames}
\label{sect:Lorentz}

In this section we write down the Lorentz transformation between the rest frame ${\cal K}$ of the plasma and the pulsar frame ${\cal K}'$ in which the plasma is streaming at speed $\beta_{\rm s}$ away from the star (positive direction). We also discuss the transformation of a 1D J\"uttner distribution between ${\cal K}$ and ${\cal K}'$.

\subsection{Lorentz transformation to the pulsar frame}

The Lorentz transformation from the unprimed frame ${\cal K}$ to the primed frame ${\cal K}'$ moving along the magnetic field at speed $\beta_{\rm s}$ applied to a wave, described by frequency $\omega$ and components $k_\parallel$ and $k_\perp$, parallel and perpendicular, respectively, to the relative velocity, gives
\be
\omega'=\gamma_{\rm s}(\omega+k_\parallel c\beta_{\rm s}),
\quad
k'_\parallel c=\gamma_{\rm s}(k_\parallel c+\omega\beta_{\rm s}),
\quad
k'_\perp=k_\perp,
\label{LT1}
\ee
with $\gamma_{\rm s}=(1-\beta_{\rm s}^2)^{-1/2}$. In terms of the variables $z=\omega/k_\parallel c$ and $\theta$ in the unprimed frame, $ \mathcal{K} $, and $z'=\omega'/k'_\parallel c$ and $\theta'$ in the primed frames, $ \mathcal{K}' $, equations (\ref{LT1}) and the inverse transforms imply
\be
z'=\frac{z+\beta_{\rm s}}{1+\beta_{\rm s} z},
\quad
z=\frac{z'-\beta_{\rm s}}{1-\beta_{\rm s} z'},
\quad
\tan\theta'=\frac{\tan\theta}{\gamma_{\rm s}(1+\beta_{\rm s} z)},
\quad
\tan\theta=\frac{\tan\theta'}{\gamma_{\rm s}(1-\beta_{\rm s} z')}.
\label{LT2}
\ee

\begin{figure}
\begin{center}
\psfragfig[width=0.8\columnwidth]{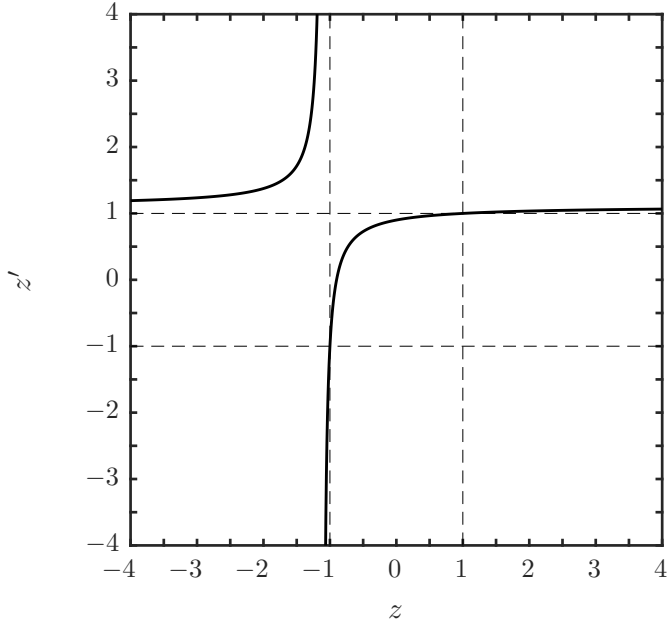}
 \caption{The relation (\ref{LT2}) between $z'$ and $z$ is plotted for $\beta_{\rm s}=0.9$; the box enclosed by the dashed lines at $z,z'=\pm1$ is the subluminal range.}
 \label{fig:zzp}  
 \end{center}
\end{figure}

The relation between $z'$ and $z$ is illustrated (for $\beta_{\rm s}=0.9$) in Figure~\ref{fig:zzp}. The relation separates into two branches. One branch includes the subluminal range, $ -1 < z < 1 $ with $ -1 < z' < 1 $, and two superluminal ranges, one where both $z,z'$ are negative, $ -1/\beta_{\rm s} < z < -1 $ with $ -\infty < z' < -1 $, and another where both $z,z'$ are positive, $ 1 < z < \infty $ with $ 1 < z' < 1/\beta_{\rm s} $. The other branch is for superluminal negative $z$ and superluminal positive $z'$, $-\infty<z<-1/\beta_{\rm s}$ with $1/\beta_{\rm s}<z'<\infty$, respectively. Assuming a source on the near side of the pulsar, only waves with $z'>0$ can reach the observer; these include not only forward propagating waves, $z>0$, in ${\cal K}$ but also backward propagating waves with $-\beta_{\rm s} < z < 0$ in ${\cal K}$, which become forward propagating waves, $z'>0$, in ${\cal K}'$. 

\subsection{Subluminal waves}

The subluminal range $-1<z<+1$ in ${\cal K}$ maps onto the subluminal range $-1<z'<+1$ in ${\cal K}'$. However, $z'$ and $z$ can have opposite signs.  The phase speed $z=0$ ($ z' = 0 $) that separates forward- and backward-propagating waves in ${\cal K}$ (in $ \mathcal{K}' $) maps onto $z'=\beta_{\rm s}$ ($ z = -\beta_{\rm s} $) in ${\cal K}'$ (in $ \mathcal{K} $). Forward-propagating waves with $0<z'<\beta_{\rm s}$ in ${\cal K}'$ correspond to backward-propagating waves $-\beta_{\rm s}<z<0$ in ${\cal K}$. However, this interpretation requires further comment. Note that the inverse of the transformation given by equation (\ref{LT1}), specifically $\omega=\gamma_{\rm s}\omega'(z'-\beta_{\rm s})/z'$ and $k_\parallel c=\gamma_{\rm s}k'_\parallel c(1-z'\beta_{\rm s})$, implies that $z$ has the opposite sign to $z'$ due to $\omega<0$, $k_\parallel>0$. The negative frequency requires interpretation. 

It is conventional to describe a wave in terms of a positive frequency, and it is always possible to do so because the dispersion equation is unchanged under $\omega,k_\parallel\to-\omega,-k_\parallel$ and hence is an even function of $z$ with positive- and negative-frequency solutions $\omega=\pm\omega_M(z)$, for some wave mode $ M $. Confusion arises because negative $z$ can be due to either $\omega$ or $k_\parallel$ being negative.  A formal way of allowing for the change in sign of the frequency under a Lorentz transformation is to distinguish between forward- and backward-propagating wave modes with dispersion relations $\omega=\omega_{M\pm}(z)>0$. One then requires that if the Lorentz transformation causes the frequency to change sign, one re-interprets this as a change in mode, from forward-propagating, $M+$, to backward-propagating, $M-$.

The mapping $z\to z'$ for $\{1-|z|,1-|z'|\}\ll1$ becomes strongly distorted for $\gamma_{\rm s}\gg1$. Important features of the wave dispersion discussed in Paper~1 occur for $\gamma_\phi=(1-z^2)^{-1/2}\gg1$, and an approximate form for the Lorentz transformation is desirable for this case. The relations (\ref{LT2}) for $\{1-|z|,\theta\} \ll1$ may be approximated by
\be
\omega' \approx 2\gamma_{\rm s}\omega,
\qquad
\gamma'_\phi \approx 2\gamma_\phi\gamma_{\rm s},
\qquad
\theta' \approx \theta/2\gamma_{\rm s},
\label{LT2p}
\ee
where we assume $\{\gamma_\phi,\gamma_{\rm s}\}\gg1$. Thus phase speeds $z\approx1$ near the speed of light, $\gamma_\phi\gg1$, in ${\cal K}$ transform into phase speeds much closer to the speed of light,  $\gamma'_\phi\approx2\gamma_{\rm s}\gamma_\phi\gg\gamma_\phi$, in ${\cal K}'$. 

The approximation (\ref{LT2p}) applies to the forward propagating, $ z > 0 $, parallel Alfv\'en (or A) mode, with the dispersion relation $z=z_{\rm A}\approx1+1/2\beta_{\rm A}^2$ transforming into $z'=z'_{\rm A}$, $z'_{\rm A}=(z_{\rm A}+\beta_{\rm s})/(1+z_{\rm A}\beta_{\rm s})\approx1+1/2\beta'^2_{\rm A}$, with $\beta'_{\rm A}\approx2\gamma_{\rm s}\beta_{\rm A}$. Similarly, the maximum frequency of the L~mode is determined by the maximum of the RPDF, at $z=z_{\rm m}$ with $\gamma_\phi=\gamma_{\rm m}$ in ${\cal K}$, and at  $z'= z'_{\rm m} = (z_{\rm m}+\beta_{\rm s})/(1+z_{\rm m}\beta_{\rm s})$, with $\gamma'_\phi=\gamma'_{\rm m}\approx2\gamma_{\rm s}\gamma_{\rm m}$ in ${\cal K}'$. The features of the wave dispersion in the small range of $0 < 1-z\ll 1$, with $ z > 0 $, discussed in Paper~1 are squeezed into an extremely narrow range (smaller by a factor $1/2\gamma_{\rm s}$) of phase speeds $0 < 1-z' \ll 1$ in ${\cal K}'$.

\subsection{Superluminal waves}
\label{sect:superluminal}

The superluminal ranges in ${\cal K}$ and ${\cal K}'$ also map into each other, but in a less obvious way than for subluminal waves. In this case changes in sign between $z$ and $z'$ occur at $(z, z') = (\pm\infty, 1/\beta_{\rm s})$, or at $ (z, z') = (-1/\beta_{\rm s}, \pm\infty)$. The frequency cannot change sign, and the introduction of $\pm$ modes is not relevant. In this case, $ k'_\parallel $ changes sign, relative to $k_\parallel$, at $ z = -1/\beta_{\rm s} $ causing backward-propagating waves in $ \mathcal{K} $ with $ z < -1/\beta_{\rm s} $ to become forward-propagating in $ \mathcal{K}' $.

In the application to pulsars, superluminal waves are relevant to oscillations that are primarily in time. Purely temporal oscillation correspond to $k_\parallel=0$, or $z=\pm\infty$, in ${\cal K}$ and to $k'_\parallel=0$, or $z'=\pm\infty$, in ${\cal K}'$, and these may be identified as the conditions for the cutoff frequencies in the two frames. However, the cutoff frequencies in the two frames are not the same (in any meaningful sense) and the relation between them is not obvious. Specifically, assuming $k_\parallel=0$ in ${\cal K}$ and $k'_\parallel=0$ in ${\cal K}'$ implies frequencies that are related by $\omega=\gamma_{\rm s}\omega'$ and $\omega'=\gamma_{\rm s}\omega$, respectively. In a pulsar plasma the only cutoff (in the radio range) is in the L~mode at $\omega=\omega_x=\omega_{\rm p}\langle1/\gamma^3\rangle^{1/2} \approx \omega_{\rm p}/\langle\gamma\rangle^{1/2}$, for $ \langle{\gamma}\rangle \gg 1 $, in ${\cal K}$, and this corresponds to $\omega'=\omega_x/\gamma_{\rm s}$ in ${\cal K}'$. On the other hand, $k'_\parallel=0$ in ${\cal K}'$ corresponds to $z=-1/\beta_{\rm s}$ in ${\cal K}$, and to a frequency $\omega=\omega_{\rm L}(-1/\beta_{\rm s})\approx\omega_{\rm L}(-1)=\omega_1$, for $ \gamma_{\rm s} \gg 1$, in ${\cal K}$, and hence to $\omega'\approx\gamma_{\rm s}\omega_1$ in ${\cal K}'$. We remark that the relation $\omega\approx\gamma_{\rm s}\omega'$ applies for nearly temporal oscillations (large $z$) in ${\cal K}$ and the relation $\omega'\approx\gamma_{\rm s}\omega$ applies for nearly temporal oscillations (large $z'$) in ${\cal K}'$. There is a rapid transition between these relations near $z=-z'$, with $|z|=|z'|=1+1/\gamma_{\rm s}$. This rapid transition near $z\lesssim-1$, $z'\gtrsim1$ is evident (for $\beta_{\rm s}=0.9$) in the upper-left branch in Figure~\ref{fig:zzp}.

\subsection{Distribution function in the pulsar frame}

The distribution $g(u)$ in the rest frame may be rewritten in the pulsar frame by noting that it is invariant under Lorentz transformations along the direction of the magnetic field. In a 4-tensor notation, let $u^\mu=(u^0,{\bi u})$ denote a 4-velocity, with $u^0=\gamma$, ${\bi u}=\gamma\beta{\bi b}$, where ${\bi b}$ is the unit vector along the magnetic field. We denote the invariant constructed from two 4-vectors $v^\mu$ and $w^\mu$ by $vw=v^0w^0-{\bi v}\cdot{\bi w}$. The 4-velocity corresponding to a system at rest is $u_0^\mu=(1,{\bf0})$ and the 4-velocity of a system moving at speed $\beta_{\rm s}$ is $u_{\rm s}^\mu=(\gamma_{\rm s},\gamma_{\rm s}\beta_{\rm s}{\bi b})$. The parameters $\gamma,\beta$ and $\gamma',\beta'$ are related by the Lorentz transformation:
\be
\gamma'=\gamma\gamma_{\rm s}(1+\beta\beta_{\rm s}), 
\quad
\beta'=\frac{\beta+\beta_{\rm s}}{1+\beta\beta_{\rm s}};
\qquad
\gamma=\gamma'\gamma_{\rm s}(1-\beta'\beta_{\rm s}),
\quad
\beta=\frac{\beta'-\beta_{\rm s}}{1-\beta'\beta_{\rm s}}.
\label{LT3}
\ee

For any distribution function in ${\cal K}$ that depends only on the energy, it is convenient to write this dependence in terms of $\gamma=u_0u$. We note the invariant $u_{\rm s}u'=u_0u$ constructed from the 4-velocity $u^\mu=(\gamma,\gamma\beta{\bi b})$ and from the 4-velocity $u'^\mu=(\gamma',\gamma'\beta'{\bi b})$. It is convenient to write the distribution function $g(u)$ in ${\cal K}$ as $g(\gamma)$, when it depends only on the energy, and to rewrite this as $g(u_0u)$. The distribution function $g'(u')$ in ${\cal K}'$ becomes $g(u_{\rm s}u')$, with $u_{\rm s}u'=\gamma_{\rm s}\gamma'(1-\beta_{\rm s}\beta')$. The normalization of $g(u)$ is fixed to the number density, $\int du\,g(u)=n$, in ${\cal K}$. The number density in ${\cal K}'$ is $n'=\int du'g'(u')$. 

As in Paper~1, we choose a 1D J\"uttner distribution, $g(u)=n\exp(-\rho\gamma)/2K_1(\rho)$ where $\rho=mc^2/T$ is the inverse temperature in units of the electron rest energy. Transforming to ${\cal K}'$ gives
\be
g'(u')=\frac{n\,e^{-\rho\gamma_{\rm s}\gamma'(1-\beta_{\rm s}\beta')}}{2K_1(\rho)},
\qquad
\int_{-\infty}^\infty du'\,g'(u')=n'={\gamma_{\rm s}}n.
\label{Juttnerp}
\ee
This result follows, for $g(-u)=g(u)$, from $du'=d(\beta'\gamma')=\gamma'^3d\beta'$, $du=\gamma^3d\beta$ and $d\beta'/d\beta=\gamma^2/\gamma'^2$ implying $du'/du=\gamma'/\gamma$, with $\gamma'$ and $\beta'$ given in terms of $\gamma$ and $\beta$ by equation (\ref{LT3}). 

\section{Streaming J\"uttner distribution}
\label{sect:beam} 

In this section we re-interpret the Lorentz-transformed J\"uttner distribution (\ref{Juttnerp}) as a streaming J\"uttner distribution and argue that this should be the preferred choice to model streaming particles in a pulsar plasma. We start by writing down a multi-beam model that consists of a sum of such transformed J\"uttner distributions with different streaming speeds. We then discuss the properties of a single such streaming distribution and compare it with a relativistically streaming Gaussian model that has been used in the pulsar literature.

\subsection{Multi-beam model}

A multi-beam model for the total distribution function of particles is assumed to consist of a number of components that are streaming relative to each other. Such a model applies in a specific frame, which we leave undefined, with each streaming speed relative to a point at rest in this frame. Let a specific distribution function, $g_\alpha(u)$, correspond to a streaming J\"uttner distribution with a streaming speed $\beta_\alpha$, inverse temperature $\rho_\alpha$ and number density $n_\alpha$.  The contribution of species $\alpha$ to the total distribution function is obtained by Lorentz transforming the J\"uttner distribution in the rest frame to the frame in which it is streaming with speed $\beta_\alpha$. Using equation (\ref{Juttnerp}), this gives
\be
g_\alpha(u)=\frac{n_\alpha}{\gamma_\alpha}\, \frac{e^{-\rho_\alpha\gamma_\alpha\gamma(1-\beta_\alpha\beta)}}{2K_1(\rho_\alpha)},
\label{Galpha}
\ee
where $n_\alpha/\gamma_\alpha$ is the number density in the rest frame of species $\alpha$. The multi-beam model corresponds to a sum of such distributions:
\be
g(u)=\sum_\alpha g_\alpha(u)
\quad\text{with}\quad
\int_{-\infty}^\infty {\rm d}u\, g(u) = n = \sum_\alpha n_\alpha.
\label{mb}
\ee
We discuss specific examples involving two such distributions in the next section.

\subsection{Relativistically streaming distributions}

In discussing choices for the distribution function of a relativistic beam in a pulsar plasma, it is helpful to start from nonrelativistic counterparts. In the absence of streaming the default choice in the nonrelativistic case is a Maxwellian distribution, $\propto\exp(-\rho\beta^2/2)$ in the notation used in this paper. The corresponding model for a beam is a distribution streaming with speed $\beta_\alpha$; this is $\propto\exp[-\rho(\beta-\beta_\alpha)^2/2]$, which is obtained by applying a Galilean transformation to the Maxwellian distribution. We discuss several different choices of relativistic (non-streaming and streaming) distributions that are generalization of the Maxwellian case. 

The standard relativistic generalization in the non-streaming case is a J\"uttner distribution, which is  obtained from the nonrelativistic Maxwellian distribution by replacing $\beta^2/2$ by $\gamma-1$, noting the expansion $\gamma=1+\beta^2/2+\ldots$ for $\beta^2\ll1$. This is equivalent to writing the Maxwellian distribution in the form $\propto\exp(-\varepsilon/T)$ and replacing the nonrelativistic energy, $\varepsilon=mc^2\beta^2/2$, by its relativistic counterpart, $\varepsilon=\gamma mc^2$. Our choice for a relativistically streaming distribution is obtained by applying a Lorentz transformation to the resulting J\"uttner distribution. A relativistically streaming J\"uttner distribution is qualitatively different from its nonrelativistic counterpart, notably in the absence of any approximate symmetry. Specifically, a streaming 1D Maxwellian distribution,  $\propto\exp[-\rho_\alpha(\beta-\beta_\alpha)^2/2]$, is symmetric about $\beta=\beta_\alpha$, but there is no such symmetry for a relativistically streaming J\"uttner distribution, $ \propto \exp[-\rho_\alpha\gamma_\alpha\gamma(1-\beta_\alpha\beta)] $. 

Another choice of relativistic generalization of a Maxwellian distribution involves replacing the 3-speed $\beta$ by the 4-speed $u=\gamma\beta$. In the absence of streaming this gives a Gaussian distribution $\propto\exp(-u^2/2u_{\rm th}^2)$, with $u_{\rm th}^2=1/\rho_\alpha$ regarded as a free parameter in the model.  This generalization applied to a streaming Maxwellian gives a streaming Gaussian, which is a favored choice in the pulsar literature \citep[e.g.,][]{LP82,AM98}:
\be
g_\alpha(u)=\frac{n_\alpha}{(2\pi)^{1/2}u_{\rm th}}
\exp\left[-\frac{(u-u_\alpha)^2}{2u_{\rm th}^2}\right].
\label{gup}
\ee
The parameter $u_{\rm th}^2$ may also be interpreted as the average $\langle(u-u_\alpha)^2\rangle$ over this distribution function. Note that the form (\ref{gup}) is obtained by two sequential replacements: including the streaming through $\beta\to\beta-\beta_\alpha$ and including relativistic effects through $ \{\beta,\beta_\alpha \} \to \{u,u_\alpha\} $. A different result is obtained if one makes these generalizations in the opposite order, cf. equation (\ref{gugs}).

We note two differences between the relativistically streaming Gaussian (\ref{gup}) and a streaming J\"uttner distribution. First, like its nonrelativistic counterpart, a relativistically streaming Gaussian is symmetric about $u=u_\alpha$, whereas there is no such symmetry for a streaming J\"uttner distribution. Second, a streaming J\"uttner distribution is related to its non-streaming counterpart by a Lorentz transformation, but there is no such relation for a relativistic Gaussian. Specifically, the Lorentz-transformed Gaussian is obtained by replacing its dependence on $u=\gamma\beta$ in terms of primed quantities using $\gamma=\gamma_\alpha\gamma'(1-\beta'\beta_\alpha)$ and $\beta=(\beta'-\beta_\alpha)/(1-\beta'\beta_\alpha)$, where a prime denotes quantities in the frame in which the distribution is streaming. The Gaussian distribution, $\propto\exp(-u^2/2u_{\rm th}^2)$, does not transform into the streaming Gaussian distribution (\ref{gup}). The Lorentz transform of any given distribution $g_\alpha(u)$ is not $g_\alpha(u-u_\alpha)$, but rather $g_\alpha(u'_\alpha)$ with $u'_\alpha=\gamma\gamma_\alpha(\beta-\beta_\alpha)$. A relativistic Gaussian in its rest frame transforms into
\begin{equation}
g_\alpha(u)\propto\exp[-\gamma^2\gamma_\alpha^2(\beta-\beta_\alpha)^2/2u_{\rm th}^2]\propto\exp[-(\gamma^2-\gamma_\alpha^2)^2/8\gamma^2\gamma_\alpha^2u_{\rm th}^2],
\label{gugs}
\end{equation}
where the final form applies for $ \{\gamma^2,\gamma_\alpha^2\} \gg 1 $. A distribution of the form (\ref{gugs}) has some similarities to the streaming J\"uttner distribution (\ref{Galpha}). However, we see no reason to prefer the distribution (\ref{gugs}) over the streaming J\"uttner distribution (\ref{Galpha}). 

\begin{figure}
\begin{center}
\psfragfig[width=0.48\columnwidth]{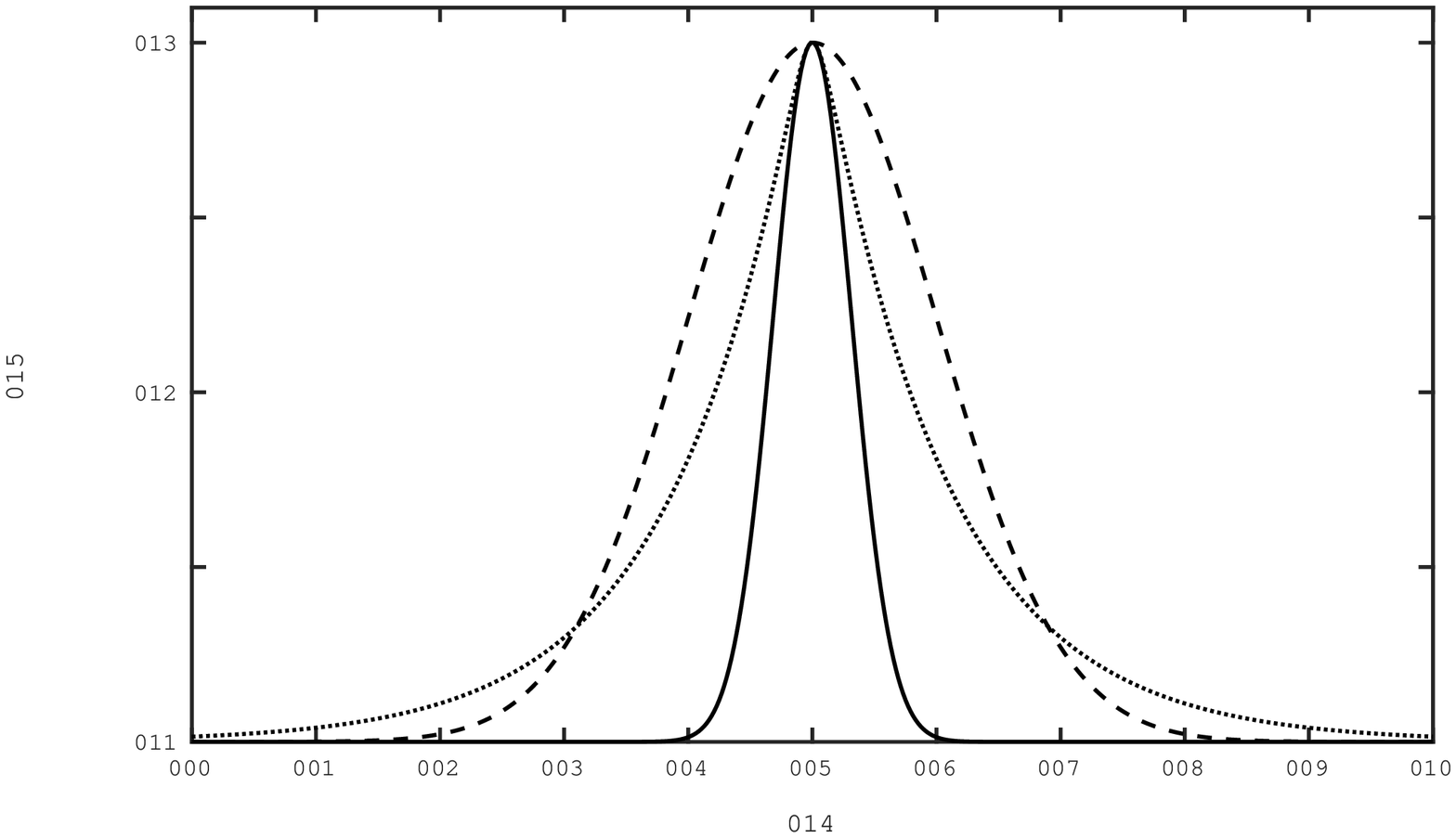}
\quad
\psfragfig[width=0.48\columnwidth]{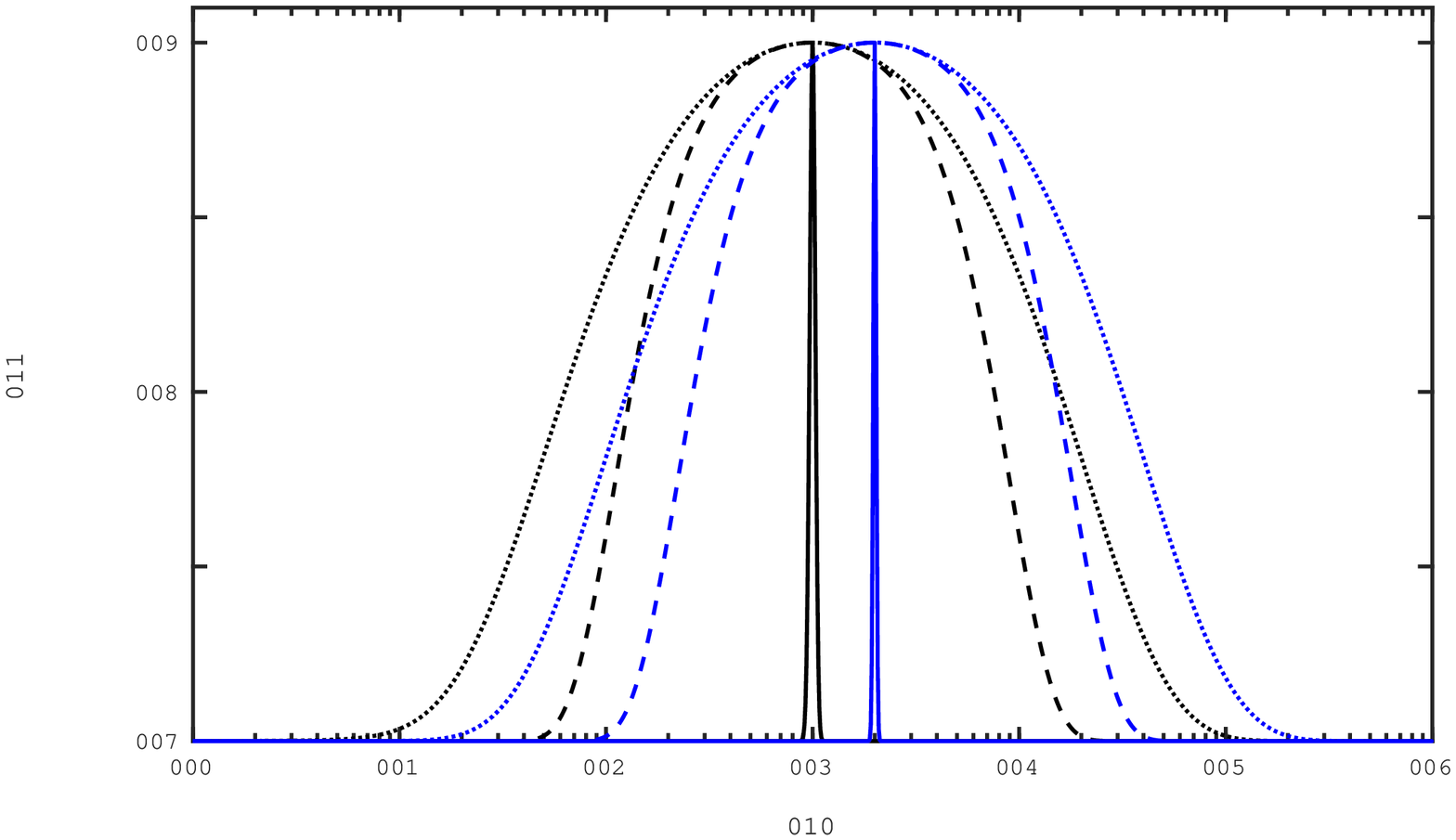}
 \caption{Plots of Gaussian distribution as given by~\eqref{gup} (solid) and as given by~\eqref{gugs} (dashed), and J\"uttner distribution as given by~\eqref{Galpha} (dotted) as a function of $u=\gamma\beta$ for $\rho_\alpha=0.1$. Left: $u_\alpha=\gamma_\alpha\beta_\alpha=0 $, with $ u_{\rm th}^2 = 1/\rho_\alpha $ (solid) and $ u_{\rm th}^2 = 1/\rho_\alpha^2 $ (dashed). Right: $ u_{\rm th}^2 = 1/\rho_\alpha $ with $u_\alpha=100 $ (black) and 200 (blue). The amplitude is in each case normalized to unity; the number density in each distribution is proportional to the area under the curve.}
 \label{fig:GJ0p1} 
 \end{center}
\end{figure}

In Figure~\ref{fig:GJ0p1} we plot the Gaussian (solid and dashed) and J\"uttner (dotted) distributions for $ \rho_\alpha = 0.1 $. In the left panel we choose $ u_\alpha = 0 $ for which the two expressions for the Gaussian distribution given by equations~\eqref{gup} and~\eqref{gugs} coincide: $ u_{\rm th}^2 = 1/\rho_\alpha $ (solid) and $ u_{\rm th}^2 = 1/\rho_\alpha^2$ (dashed); and the J\"uttner distribution~\eqref{Galpha} is given by the dotted curve. Comparison of the three cases shows that for small $|u|$ the width of the J\"uttner distribution is intermediate between a Gaussian with $u_{\rm th\alpha}^2=1/\rho_\alpha$ and a Gaussian with $u_{\rm th\alpha}^2=1/\rho_\alpha^2$, with the J\"uttner distribution having much broader wings at larger $|u|$. The number density is proportional to the area under the curve, $\propto\gamma_\alpha$. The change when streaming is included is shown in the right panel for $ u_{\rm th}^2 = 1/\rho_\alpha $ with $ u_\alpha = 100 $ (black curves) and 200 (blue curves). The solid curves show plots of the Gaussian distribution as given by equation~\eqref{gup} and the dashed curves show the form given by equation~\eqref{gugs}. The corresponding plots for the J\"uttner distribution are given by the dotted curves. It is clear that the Lorentz-transformed Gaussian distribution~\eqref{gugs} is much broader, with its width increasing as $ u_\alpha $ increases, whereas the width of the shifted Gaussian~\eqref{gup} is independent of $ u_\alpha $. Below the peak at $u=u_\alpha$, the positive slope of the J\"uttner distribution is much smaller than for either Gaussian, and above the peak the J\"uttner distribution decreases much more slowly with $u$ than for either Gaussian. The width of the Lorentz-transformed Gaussian remains comparable to that of the J\"uttner distribution when plotted as a function of the logarithm of $ u = \gamma\beta $.

The streaming Gaussian distribution (\ref{gup}) is a poor approximation to a streaming J\"uttner distribution  for $ \rho_\alpha \ll 1 $. In particular the slope of the distribution, ${\rm d}g_\alpha(u)/{\rm d}u$, for either Gaussian distribution is a poor approximation to the slope for the J\"uttner distribution. This slope is directly relevant to a beam-driven instability, suggesting that the growth rate for a J\"uttner distribution is poorly approximated by a streaming Gaussian model.

We suggest that the choice of a relativistically streaming Gaussian distribution (\ref{gup}) is made primarily for mathematical convenience. The choice (\ref{gup}) applies only in a specific frame, in the sense that it does not retain its form under a Lorentz transformation. 

Another choice of (non-streaming) distribution function, made primarily for mathematical convenience, is of the form  $g(u)\propto(u_1^2-u^2)^N$, for $u^2<u_1^2$ and $g(u)=0$ for $u^2>u_1^2$, where $u_1$ is a constant, with the cases $n=0,1,2$ referred to as water-bag \citep{AB86}, hard-bell and soft-bell \citep{GMG98,MG99} distributions, respectively. Applying a Lorentz transformation gives $g'(u')=g(u)$ with $u=\gamma_\alpha\gamma'(\beta'-\beta_\alpha)$. The range $-u_1<u<u_1$ in ${\cal K}$ transforms into $u'_{1-}<u'<u'_{1+}$ in ${\cal K}'$, with $u'_{1\pm}=\gamma_\alpha\gamma_1(\pm\beta_1+\beta_\alpha)$ implying that the range $2u_1$ in ${\cal K}$ broadens to $u'_{1+}-u'_{1-}=\gamma_\alpha2u_1$ in ${\cal K}'$. In the highly relativistic case, $\gamma_\alpha\gg\gamma_1\gg1$, these limits become $u'_{1-}\approx-(\gamma_1^2-\gamma_\alpha^2)/2\gamma_1\gamma_\alpha$, $u'_{1+}\approx\gamma_\alpha\gamma_1$. This much larger range in ${\cal K}'$ implies that the number density in ${\cal K}'$ is higher than that in ${\cal K}$, by the same factor, $n'/n=\gamma_\alpha$, as is obvious in the case $N=0$, where $g'(u')=g(u)$ is a constant, and is easily shown for $N>0$. The parameter $\gamma_1$ may be interpreted in terms of $\langle\gamma\rangle_\alpha$, with $\gamma_1=2\langle\gamma\rangle_\alpha$ for $N=0$ and $\gamma_1=8\langle\gamma\rangle_\alpha/3$ for $N=1$.

We adopt the view that the default choice for a relativistic distribution is a J\"uttner distribution in the rest frame of the plasma, and that the default choice for a relativistically streaming distribution is that obtained by applying a Lorentz transformation to the distribution function in the rest frame. The fact that the resulting streaming distribution is very much broader than the rest-frame distribution is a characteristic feature, which applies to but is not restricted to a J\"uttner distribution.

\subsection{Examples of relativistically streaming distributions}

\begin{figure}
\begin{center}
\psfragfig[width=0.48\columnwidth]{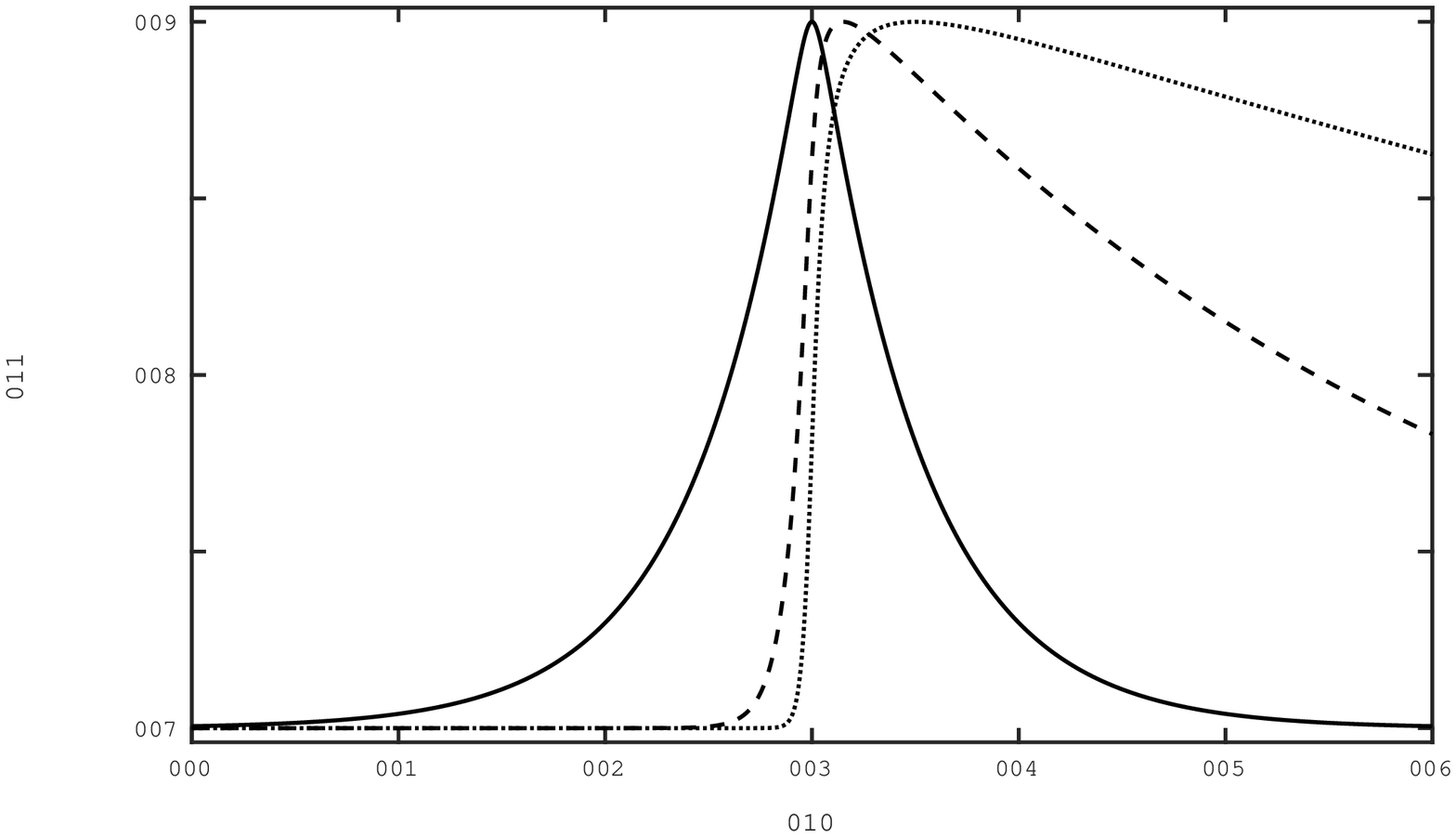}
\quad
\psfragfig[width=0.48\columnwidth]{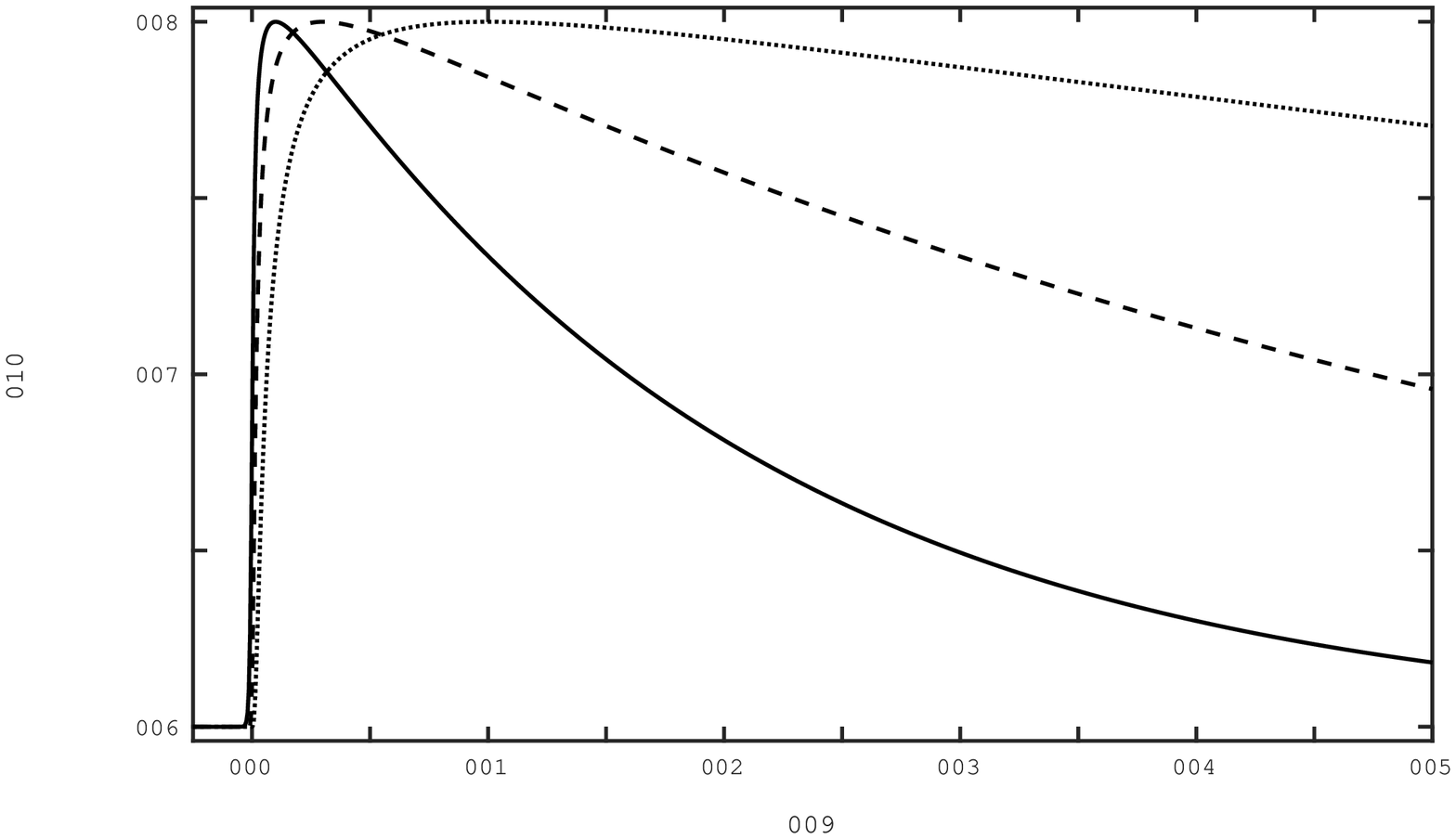}
 \caption{A non-streaming and streaming J\"uttner distributions are plotted as a function of $u=\gamma\beta$ for $\rho_\alpha=0.1$. Left: $u_\alpha=\gamma_\alpha\beta_\alpha=0 $ (solid), 3 (dashed), 10 (dotted). Right: $u_\alpha=10 $ (solid), 30 (dashed), 100 (dotted). The amplitude is in each case normalized to unity; the number density in each distribution is proportional to the area under the curve.}
 \label{fig:J0p1} 
 \end{center}
\end{figure}

In Figure~\ref{fig:J0p1} we plot the distribution function (\ref{Galpha}) for $\rho_\alpha=0.1$, and for several values of $u_\alpha=\gamma_\alpha\beta_\alpha$. On the left panel is shown a non-streaming distribution, $\beta_\alpha=0$ (solid), and two streaming distributions, $\gamma_\alpha\beta_\alpha=3$ (dashed) and $\gamma_\alpha\beta_\alpha=10$ (dotted).  The non-streaming distribution is symmetric about the origin, $u=0$; a slight asymmetry develops for a small streaming speed, and for $\gamma_\alpha\beta_\alpha \approx 1/\rho_\alpha \approx \langle\gamma\rangle_\alpha$, for $ \rho_\alpha \ll 1 $, the asymmetry is substantial. In the case $u_\alpha\approx\langle\gamma\rangle_\alpha\approx10$ the distribution function is almost negligible for $u < 0 $, and increases with increasing $u>0$ to a maximum near $u=u_\alpha\approx10$, and then decreases slowly for $u\gg u_\alpha$. On the right panel in Figure~\ref{fig:J0p1} we show the cases $u_\alpha=10 $ (solid), 30 (dashed), 100 (dotted) on a larger scale. In each case the distribution function has a maximum at $u=u_\alpha$. Note that the normalization in Figure~\ref{fig:J0p1} is chosen to show the relative shapes of the distributions: each is normalized so that its maximum is unity. The number density in each case is proportional to the area under the curve, which is $\propto\gamma_\alpha$ for a streaming J\"uttner distribution; with normalization to a fixed number density the maxima would be $\propto1/\gamma_\alpha$.\footnote{This result applies to any distribution that satisfies $g(-u)=g(u)$ in its rest frame.}

\subsection{Highly relativistically streaming J\"uttner distribution} 

An analytic approximation to a highly relativistically streaming J\"uttner distribution, with $\rho_\alpha\ll1$ and $\rho_\alpha\gamma_\alpha\gg1$, may be found by expanding the exponential factor in equation~\eqref{Galpha} in powers of $ 1/\gamma_\alpha \ll 1 $ and $ 1/\gamma\ll 1$:
\be
\exp\left[-\rho_\alpha\gamma_\alpha\gamma(1-\beta_\alpha\beta)\right]\approx
\begin{dcases}
\exp\left[-\rho_\alpha{\frac{(\gamma-\gamma_\alpha)^2}{2\gamma_\alpha\gamma}}-\rho_\alpha\right]
& \text{for} \quad \beta>0,\\
0& \text{for} \quad \beta<0.
\end{dcases}
\label{gup1}
\ee
This gives
\be
g_\alpha(u)=\frac{1}{2}n_\alpha\rho_\alpha
\exp\left[-\rho_\alpha\frac{(\gamma-\gamma_\alpha)^2}{2\gamma\gamma_\alpha}\right]\,H(u),
\label{gup2}
\ee
where $H(u)=1$ for $u>0$ and $H(u)=0$ for $u<0$ is the step function, and we use $ 1/K_1(\rho_\alpha) = \rho_\alpha + \mathcal{O}(\rho_\alpha^3) $ for $ \rho_\alpha \ll 1 $.

\begin{figure}
\begin{center}
\psfragfig[width=0.60\columnwidth]{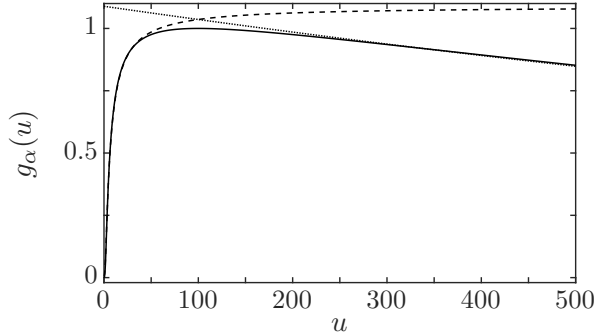}
 \caption{Comparison of the approximate forms~\eqref{gup3} and the exact form~\eqref{Galpha} of a J\"uttner distribution with $\rho_\alpha=0.1$ and $u_\alpha=100$. The solid curve is the exact distribution, the dashed curve is the approximation for $ \gamma \ll \gamma_\alpha $, and the dotted curve is the approximation for $ \gamma \gg \gamma_\alpha $.}
 \label{fig:gapprox}  
 \end{center}
\end{figure}

The approximation (\ref{gup1}) gives approximate forms for the highly relativistically streaming J\"uttner distribution function below and above the peak of the distribution at $\gamma=\gamma_\alpha$:
\be
g_\alpha(u)\approx \frac{1}{2}n_\alpha\rho_\alpha
\begin{dcases}
\exp\left(-\rho_\alpha\gamma_\alpha/2\gamma\right)
&\text{for}\quad \gamma\ll\gamma_\alpha,\\
\exp\left(-\rho_\alpha\gamma/2\gamma_\alpha\right)
&\text{for}\quad \gamma\gg\gamma_\alpha.
\end{dcases}
\label{gup3}
\ee
The exact form and the two approximate forms (\ref{gup3}) are compared in Figure~\ref{fig:gapprox}. The approximation for $\gamma\ll\gamma_\alpha$ implies a positive slope ${\rm d}g_\alpha(u)/{\rm d}u\approx(\rho_\alpha\gamma_\alpha/2\gamma^2)g_\alpha(u)$. The approximation for $\gamma\gg\gamma_\alpha$ implies that a relativistically streaming J\"uttner distribution asymptotes to the same form as a non-streaming J\"uttner distribution, with $\rho_\alpha$ replaced by $\rho_\alpha/2\gamma_\alpha$ in the exponent.

For comparison we consider the same highly relativistic approximation to the Lorentz-transformed Gaussian distribution (\ref{gugs}). In place of equation (\ref{gup3}), this gives
\be
g_\alpha(u)\propto
\begin{dcases}
\exp\left(-\gamma_\alpha^2/8\gamma^2u_{\rm th}^2\right)
&\text{for}\quad \gamma^2\ll\gamma_\alpha^2,\\
\exp\left(-\gamma^2/8\gamma_\alpha^2u_{\rm th}^2\right)
&\text{for}\quad \gamma^2\gg\gamma_\alpha^2.
\end{dcases}
\label{gugsur}
\ee
The distribution (\ref{gugsur}) is very much broader than the conventional form  (\ref{gup}) for a relativistically streaming Gaussian distribution, as is evident by the way in which they fall off for $\gamma^2\gg\gamma_\alpha^2$: specifically $\propto\exp(-\gamma^2/2u_{\rm th}^2)$ and $\propto\exp(-\gamma^2/8\gamma_\alpha^2u_{\rm th}^2)$, respectively.

\section{``Separation'' of relatively streaming distributions}
\label{sect:separation}

In the familiar bump-in-tail instability, in which Langmuir waves grow due to a beam of fast electrons in a nonrelativistic plasma, growth requires a minimum in the total distribution function between the thermal background and the fast beam. In this section we discuss the generalization of this ``separation'' condition to the relativistic case for J\"uttner distributions. We first estimate the condition for separation between two counter-streaming distributions.

\subsection{Counter-streaming distributions}
\label{sect:counter}

An  idealized counter-streaming distribution consists of two streaming J\"uttner distributions, $\alpha=\pm$, with ${\bar n}_\pm={\bar n}/2$, $|\beta_\pm|={\bar\beta}$, and the same temperature $\rho_\pm=\rho\approx1/\langle\gamma\rangle\ll1$.  The resulting distribution function is
\be
g_{\rm cs}(u)=\frac{{\bar n}F(u)}{{2\bar\gamma}K_1(\rho)},
\qquad
F(u)=\exp[-\rho{\bar\gamma}\gamma(1-\beta{\bar\beta})]
+\exp[-\rho{\bar\gamma}\gamma(1+\beta{\bar\beta})].
\label{gucs}
\ee
We first discuss how the distribution changes as the speed ${\bar u}={\bar\gamma}{\bar\beta}$ increases from zero to ${\bar\gamma}\gtrsim1/\rho\gg1$. We then transform to the frame where one of the distributions is at rest and consider the highly relativistic case.

\begin{figure}
\begin{center}
\psfragfig[width=0.48\columnwidth]{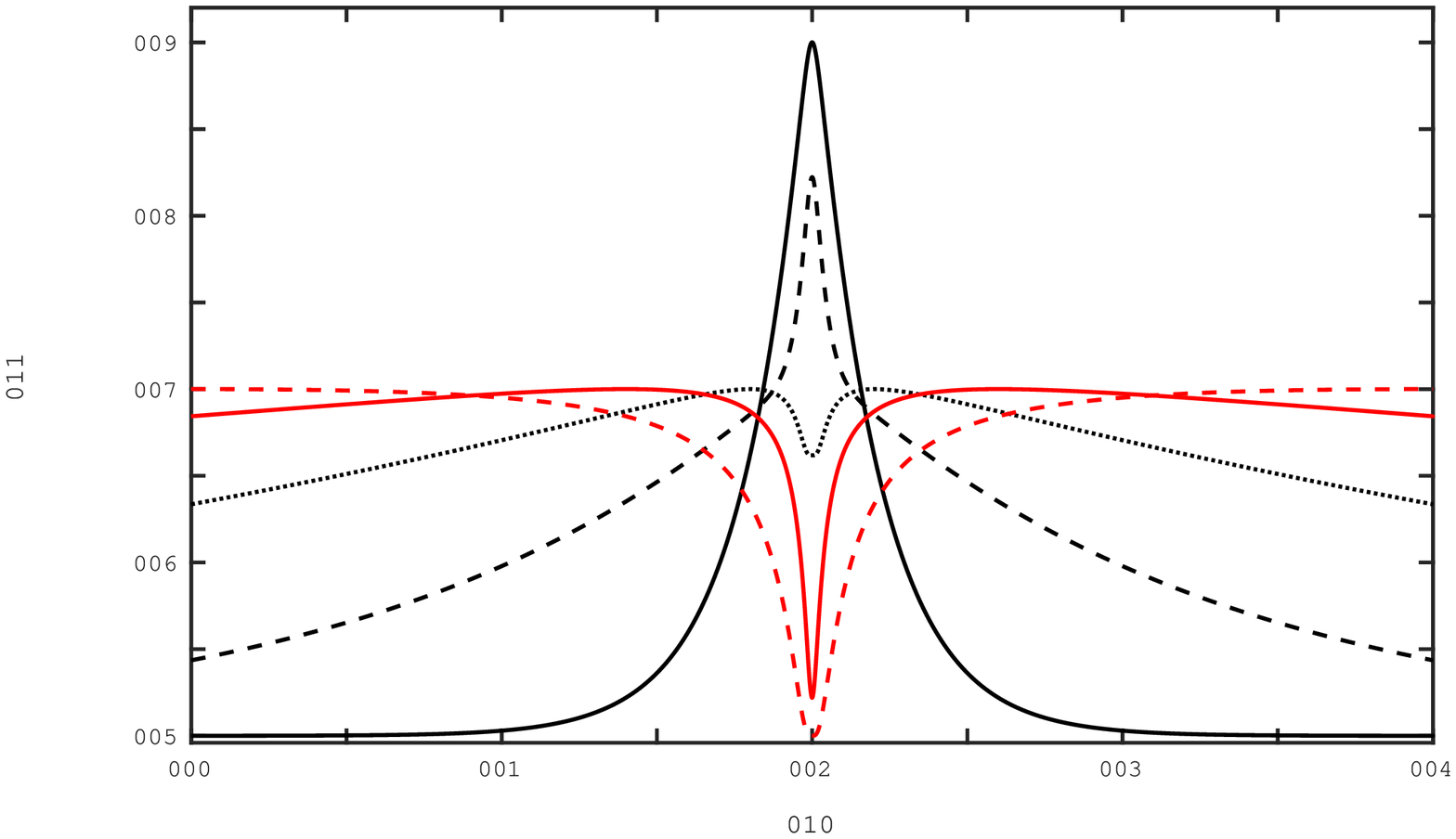}
\psfragfig[width=0.48\columnwidth]{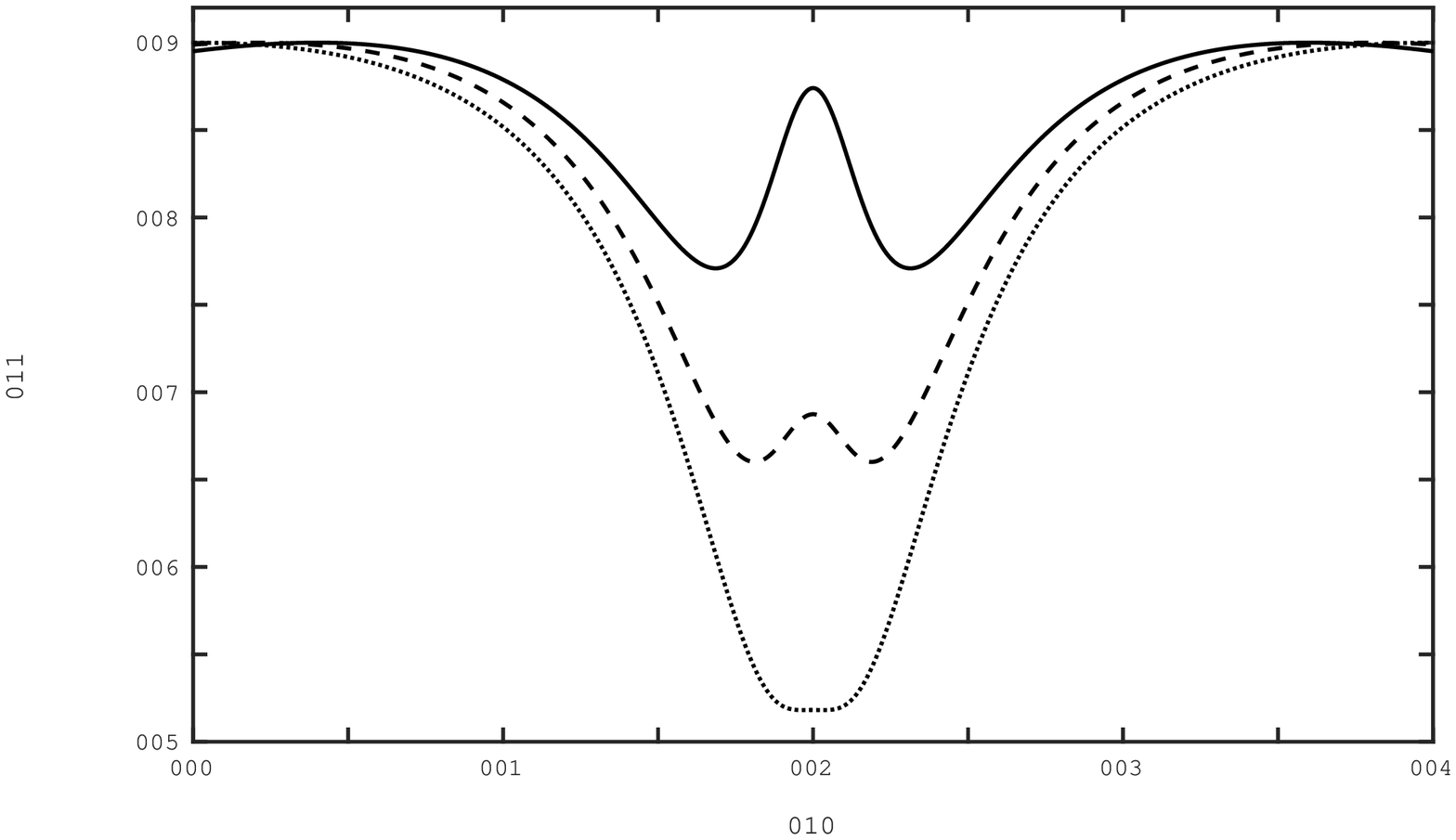}
 \caption{The sum of two counter-streaming J\"uttner distributions is plotted as a function of $u=\gamma\beta$ for $\rho=0.1$. Left: for ${\bar u}={\bar\gamma}{\bar\beta}=0 $ (black solid), 3 (black dashed), 10 (black dotted), 30 (red solid), 100 (red dashed). Right: for ${\bar u}= 8 $ (solid), 9 (dashed), 10 (dotted) showing the disappearance of the peak at $u=0$. The normalization of both distributions is chosen such that its maximum, at $u=\pm{\bar u}$, is unity. }
  \label{fig:gcs} 
 \end{center}
\end{figure}

For ${\bar u}=0$ the two distributions are identical, and their sum is a single J\"uttner distribution, corresponding to the solid black curve on the left in Figure~\ref{fig:gcs}. As shown in Figure~\ref{fig:gcs}, the curves move apart with increasing ${\bar u}$, becoming almost completely separated for ${\bar u}\gg10$. This ``separation condition'' is important in estimating the conditions under which the combined distribution can be interpreted as a beam propagating through a background distribution.
The separation condition was discussed by \citet{LSS10}, who considered the 3D counterpart, but this difference is unimportant here. Separation occurs for $ \bar{\gamma}\bar{\beta}^2 > 1/\rho $, as shown on the right in Figure~\ref{fig:gcs}. For $ \{\bar{u}, \langle\gamma\rangle\} \gg 1 $ we may write this separation condition as $ \bar{u}/\langle\gamma\rangle \gtrsim 1 $.

\subsubsection{Transformation to rest frame of one beam}

The properties of the counter-streaming distribution are useful in discussing the weak-beam model. The idea is that by transforming to the frame in which the backward propagating distribution is at rest, the backward propagating distribution is re-interpreted as the background distribution, with the forward propagating distribution being regarded as the beam. The weak-beam case follows by multiplying the latter distribution by the ratio of the beam to background densities. The relative speed between the two distributions becomes the beam speed 
\be
\beta_{\rm b}=\frac{2{\bar\beta}}{1+\bar\beta^2}, 
\qquad
\gamma_{\rm b}=(1+\bar\beta^2){\bar\gamma}^2,
\qquad
u_{\rm b}=\gamma_{\rm b}\beta_{\rm b}=2{\bar\beta}{\bar\gamma}^2.
\label{ub1}
\ee

Let a quantity in the frame in which the backward propagating beam is at rest be  denoted by a prime. Then in equation (\ref{gucs}) one has
\be
\gamma{\bar\gamma}(1+\beta{\bar\beta})=\gamma',
\qquad
\gamma{\bar\gamma}(1-\beta{\bar\beta})=\gamma'\gamma_{\rm b}(1-\beta'\beta_{\rm b}).
\label{ub2}
\ee
Let $n_0$ be the number density of either beam in the rest frame of that beam. Using the fact that $g'_{\rm cs}(u')=g_{\rm cs}(u)$ is an invariant, in the primed frame equation (\ref{gucs}) becomes
\be
g'_{\rm cs}(u')=\frac{n_0}{2K_1(\rho)}
\left\{\exp(-\rho\gamma')
+\frac{n_{\rm b}}{\gamma_{\rm b}n_0}\exp[-\rho\gamma'\gamma_{\rm b}(1-\beta'\beta_{\rm b})]
\right\},
\label{gucsp}
\ee
with $n_{\rm b}=\gamma_{\rm b}n_0$ in this case. Equation (\ref{gucsp}), with primes omitted, becomes a weak-beam model for $ n_{\rm b}/\gamma_{\rm b}n_0\ll1$. 

\subsection{Separation condition}

The condition ${\bar u} \approx \bar{\gamma}\gtrsim\langle\gamma\rangle$ for two identical counter-streaming J\"uttner distributions to become well separated, transforms into $\gamma_{\rm b}\gtrsim2\langle\gamma\rangle^2$ in the frame in which one of the beams is at rest. The Lorentz transformation to the new frame makes this separation condition appear to be more extreme than in the counter-streaming frame. This separation condition is a direct result of the Lorentz transformation, and is not specific to J\"uttner distributions, as may be seen by considering counter-streaming distributions that are Gaussian in their respective rest frames.

In contrast, for counter-streaming Gaussian distributions of the form given by equation (\ref{gup}), with $\alpha=\pm$, $u_\pm=\pm{\bar u}$, $n_\pm={\bar n}/2$ and the same $u_{\rm th}$, the separation condition is closely analogous to that for the corresponding nonrelativistic counterpart, in which a Gaussian distribution is equivalent to a Maxwellian distribution. The two distributions become well separated when the streaming speeds exceed the thermal spreads, corresponding to ${\bar u}\gtrsim u_{\rm th}$. This condition transforms into $\gamma_{\rm b}\gtrsim2u_{\rm th}^2$, which is equivalent to $\gamma_{\rm b}\gtrsim2\langle\gamma\rangle^2$ for J\"uttner distributions.

\subsection{Weak-beam model}

In a weak-beam model there are only two components, which we denote by $\alpha=0,b$, where $\alpha=0$ refers to the background and $\alpha=b$ refers to the beam. The frame of interest is identified as the rest frame of the background in this case. The distribution function is then $ g(u) = g_0(u) + g_{\rm b}(u) $,
\be
g(u)=n_0\left[\frac{e^{-\rho_0\gamma}}{2K_1(\rho_0)}
+\varepsilon_n\frac{e^{-\rho_{\rm b}\gamma_{\rm b}\gamma(1-\beta\beta_{\rm b})}}{2K_1(\rho_{\rm b})}
\right],
\label{guwb}
\ee
with $ \varepsilon_n = n_{\rm b}/\gamma_{\rm b}n_0\ll1$, where the first term is $ g_0(u) $ and the second term is $ g_{\rm b}(u) $. For $\rho_0=\rho_{\rm b}=\rho$, this result also follows from equation (\ref{gucsp}) by omitting the primes and allowing arbitrary $n_{\rm b}/\gamma_{\rm b}n_0\ll1$.

A conventional approach to treating wave dispersion in this case is based on an expansion in $\varepsilon_n \ll1$. To zeroth order the beam is ignored, such that the wave dispersion is determined by the background plasma alone. To first order the beam contributes a correction to the frequency, which includes both imaginary and real parts, with the former determining the growth rate of any beam-driven instability. In the case of a maser instability, due to negative absorption, growth requires ${\rm d}g(u)/{\rm d}u>0$ at the resonant frequency, determined by $u=\gamma\beta=\gamma_\phi z$.

\begin{figure}
\begin{center}
\psfragfig[width=0.48\columnwidth]{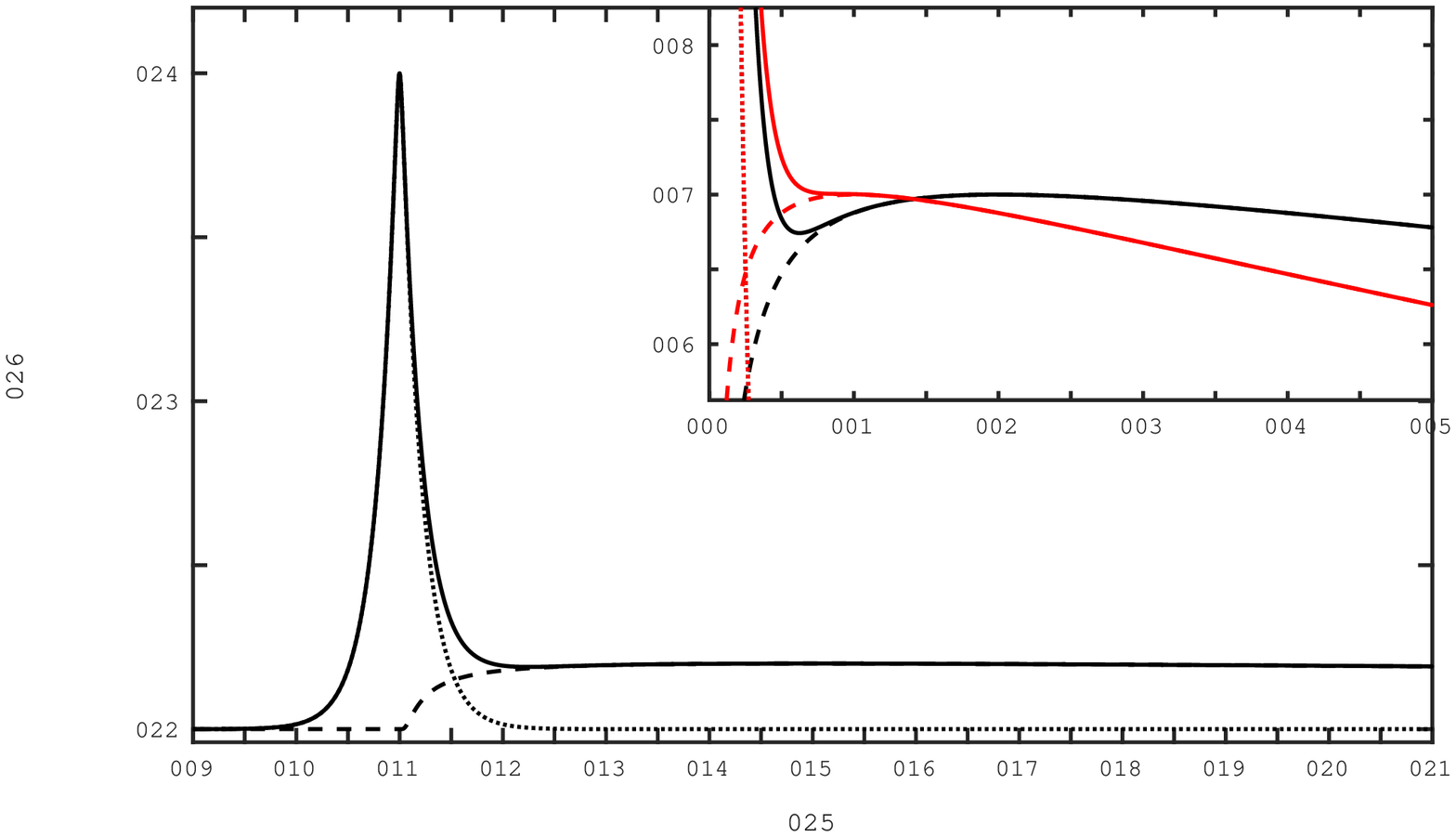}
\psfragfig[width=0.48\columnwidth]{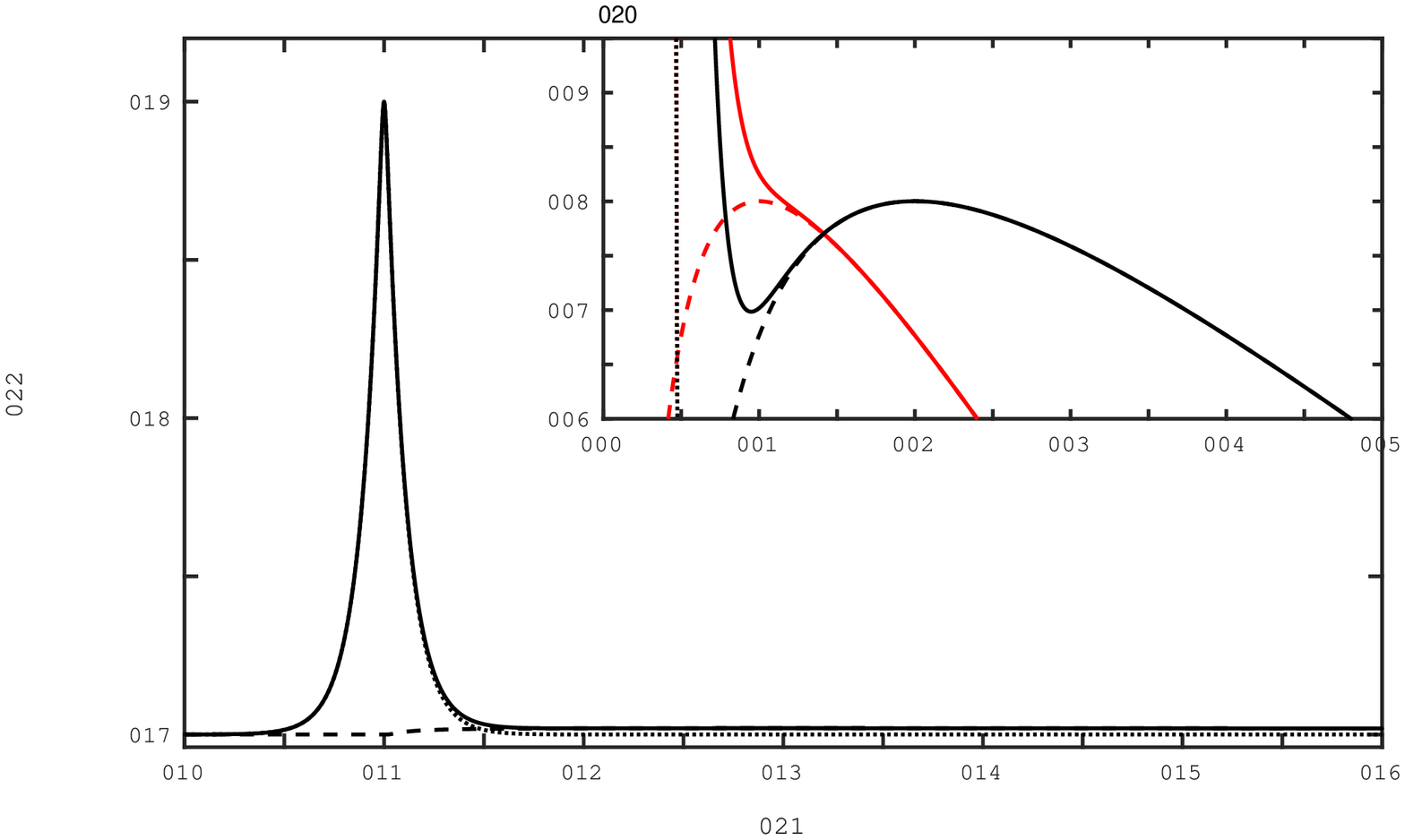}
 \caption{Weak beam distribution functions with $\rho=0.1$ for (left) $\varepsilon_n=0.1$ and (right) $\varepsilon_n=0.01$, and for $u_{\rm b}=200$ (black curves). In the inset we show a zoomed version of each plot and also include a plot for $ u_{\rm b} = 100$ (red curves). The solid curve corresponds to $ g(u) = g_0(u) + g_{\rm b}(u) $, the dashed and dotted curves are the background and beam distributions, respectively.}
  \label{fig:wb} 
 \end{center}
\end{figure}

In Figure~\ref{fig:wb} we plot the weak-beam distribution function (\ref{guwb}) for $\rho=0.1$, for two values of $\varepsilon_n = 0.1$ on the left and $0.01$ on the right, and for $u_{\rm b}=100,200$. For $\varepsilon_n=0.1$ the minimum and maximum (at $u=u_{\rm b}$) in $g(u)$ that would be present in the absence of the background have almost disappeared for $u_{\rm b}=100$, but are still present for $u_{\rm b}=200$. For $\varepsilon_n=0.01$ the minimum and maximum for $u_{\rm b}=100$ are nearly smoothed out. 

\begin{figure}
\centering
\psfragfig[width=1.0\columnwidth]{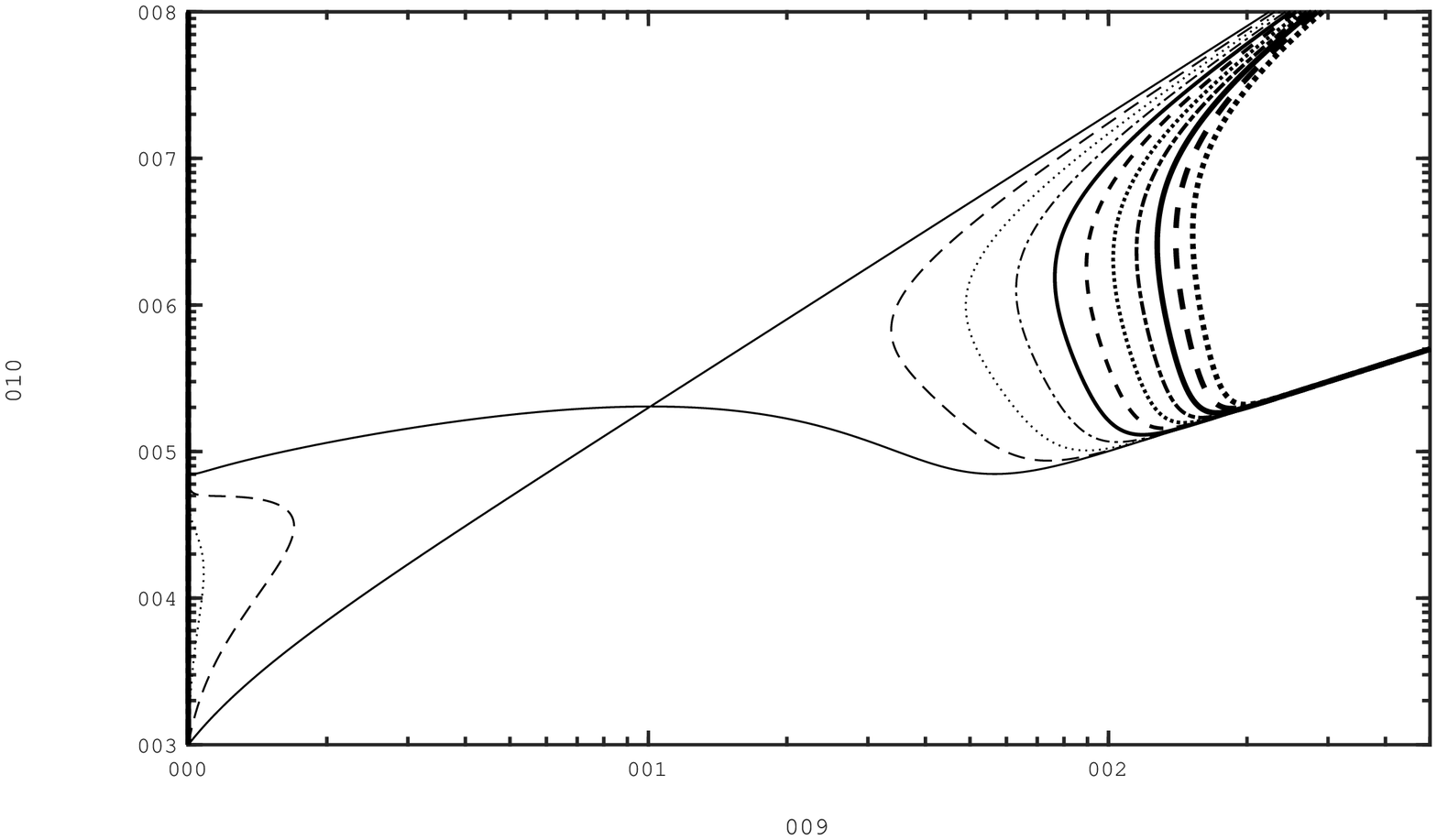}\vspace{-3mm}
\psfragfig[width=1.0\columnwidth]{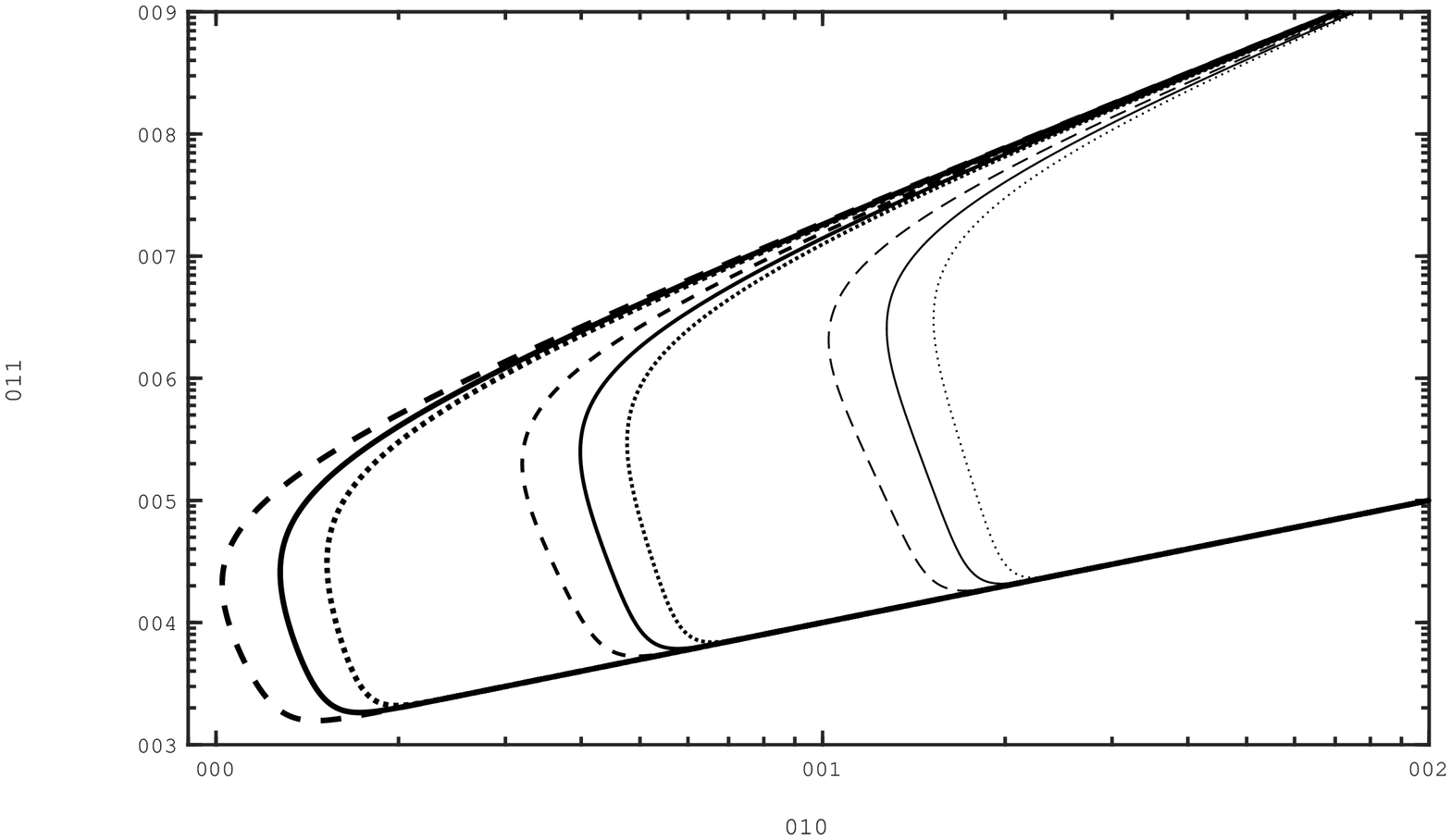}
 \caption{Contour plots of $ dg(u)/d\beta\big|_{\beta = z} = 0 $ over $ \gamma_\phi $ and $ \gamma_{\rm b} $. TOP: for $ \rho_0 = 0.1 $, $ \varepsilon_\rho = 1 $, with $ \varepsilon_n $ sampled logrithmically from $ 10^{-5} $ (thick dotted) to $ 1 $ (thin solid). BOTTOM: for $ \varepsilon_\rho = 1 $, $ \rho_0 = 0.1 $ (thick), 0.032 (medium), 0.01 (thin) with $ \varepsilon_n = 10^{-5} $ (dotted), $ 10^{-4} $ (solid) and $ 10^{-3} $ (dashed).}
 \label{fig:dgdebeta}  
\end{figure}

For $ \varepsilon_n \neq 1 $ the separation condition $\gamma_{\rm b}\gtrsim2\langle\gamma\rangle_0^2$ is modified, where $ \langle{\cdots}\rangle_0 $ indicates average over the background distribution. In Figure~\ref{fig:dgdebeta} we show contour plots of $ dg(u)/d\beta\big|_{\beta = z} = 0 $ over $ \gamma_\phi $ and $ \gamma_{\rm b} $ for various values of $ \rho_0 $ and $ \varepsilon_n $. TOP: for $ \rho_0 = 0.1 $, $ \varepsilon_\rho = \rho_{\rm b}/\rho_0 = 1 $ and $ \varepsilon_n $ sampled logarithmically from $ 10^{-5} $ (thick dotted) to $ 1 $ (thin solid). The thin solid curve corresponding to $ \varepsilon_n = 1$ is a separatrix with the solution $ \gamma_\phi \approx 2 \langle\gamma\rangle_0^2 $ denoting a saddle point at $ (\gamma_\phi, \gamma_{\rm b}) \approx (10, 200) $, for $ \langle\gamma\rangle_0 \approx 1/\rho_0 = 10 $. Solutions of $ dg(u)/d\beta\big|_{\beta = z} = 0 $ lie in the two region shown when $ \varepsilon_n < 1 $ and lie outside these regions for $ \varepsilon_n > 1 $ (which is not of interest here). For $ 3\rho_0 \lesssim \varepsilon_n \leq 1 $ the derivative of the distribution function has three peaks (and two troughs) for each $ \gamma_{\rm b} $ over the range $ \gamma_{\rm b, min} \lesssim \gamma_{\rm b} \lesssim 2\langle\gamma\rangle_0^2 $, where
\begin{equation}\label{eq:gammabmin}
    \gamma_{\rm b, min} \approx
    7.8(1/\varepsilon_n\varepsilon_\rho\varepsilon_K)^{0.076}(1/\rho_0)^{1.07}
\end{equation}
is obtained numerically, with $ \varepsilon_K = K_1(\rho_0)/K_1(\rho_{\rm b}) \approx \varepsilon_\rho $ for $ \{\rho_0, \rho_{\rm b}\} \ll 1 $. For $ \varepsilon_n \lesssim 3\rho_0 $, for each $ \gamma_{\rm b} $ the derivative always has a maximum at $ \gamma_\phi \approx 1 $ (near the peak of the background) which lies on the vertical axis in the top panel of Figure~\ref{fig:dgdebeta}, and for $ \gamma_{\rm b} \gtrsim \gamma_{\rm b, min} $ a second peak is at $ \gamma_\phi \approx \gamma_{\rm b} $ (near the peak of the beam) which follows the lower branch of the separatrix: $ \gamma_{\rm b} \approx \gamma_\phi $. The solution corresponding to the trough between these two peaks deviates from the lower branch of the separatrix at $ \gamma_{\rm b} \approx \gamma_{\rm b, min} $, rises sharply and then asymptotically approaches the upper branch of the separatrix: $ \gamma_{\rm b} \approx (2/\varepsilon_\rho)\gamma_\phi^2 $. BOTTOM: for $ \varepsilon_\rho = 1 $, $ \rho_0 = 0.1 $ (thick), 0.032 (medium), 0.01 (thin) with $ \varepsilon_n = 10^{-5} $ (dotted), $ 10^{-4} $ (solid) and $ 10^{-3} $ (dashed). The asymptotic behaviour discussed is more readily evident here where we increase the range of $ \gamma_{\rm b} $ and focus on the peak at $ \gamma_\phi \approx \gamma_{\rm b} $ and the trough of $ dg(u)/d\beta\big|_{\beta = z}$. The lower branch of the separatrix and the contour of the trough are given by
\begin{equation}\label{eq:gamma12}
    \gamma_{\rm b} 
        = \gamma_\phi + \frac{\gamma_\phi^3e^{-\rho_0\gamma_\phi}}{\varepsilon_n\varepsilon_\rho\varepsilon_K}
        \approx \gamma_\phi
    \quad\text{and}\quad
   1 - \frac{\varepsilon_n\varepsilon_\rho\varepsilon_K\gamma_{\rm b}}{2\gamma_\phi^2}\exp\left[\frac{\gamma_\phi}{\langle\gamma\rangle_0}\left(1 - \frac{\varepsilon_\rho\gamma_{\rm b}}{2\gamma_\phi^2}\right)\right] = 0,
\end{equation}
respectively, where we assume $ \{\gamma_\phi, \gamma_{\rm b}\} \gg 1 $. The background and the beam may be considered as separated if for $ \varepsilon_n \ll 1 $ we have $ \gamma_{\rm b} \gtrsim \gamma_{\rm b, min} $ with $ \gamma_{\rm b, min} $ given by~\eqref{eq:gammabmin}.

\section{Lorentz transformation of the dielectric tensor}
\label{app:LT}

The dielectric tensor in the unprimed and primed frames may be written as
\be
K_{ij}(\omega,{\bi k})=\delta_{ij}+\frac{\Pi_{ij}(\omega,{\bi k})}{\omega^2},
\qquad
K'_{i'j'}(\omega',{\bi k}')=\delta_{i'j'}+\frac{\Pi_{i'j'}(\omega',{\bi k}')}{\omega'^2},
\label{KKp}
\ee
which has the advantage that the components $\Pi_{ij}(\omega,{\bi k})$ are the space components of a 4-tensor\footnote{This is the reason why there is no prime on the kernel symbol $\Pi$ in the primed frame.} \citep{Me73,M08,M13} and hence have known properties under a Lorentz transformation. 

\subsection{Transformation of $\Pi_{ij}$}

The specific forms for the nonzero components of $\Pi_{ij}$ found in Paper~1 are
\begin{equation}
\begin{gathered}
    \Pi_{11} = \Pi_{22} = \frac{\omega_{\rm p}^2}{\Omega_{\rm e}^2}\frac{\omega^2}{z^2}\left\langle\gamma(z-\beta)^2\right\rangle,\quad
    \Pi_{13}=\Pi_{31}=\frac{\omega_{\rm p}^2}{\Omega_{\rm e}^2}\frac{\omega^2\tan\theta}{z^2}\left\langle\gamma\beta(z-\beta)\right\rangle,\\\ms
    \Pi_{33}=-\omega_{\rm p}^2z^2\left\langle\frac{1}{\gamma^3(z-\beta)^2}\right\rangle + \frac{\omega_{\rm p}^2}{\Omega_{\rm e}^2}\frac{\omega^2\tan^2\theta}{z^2}\left\langle\gamma\beta^2\right\rangle.
\label{Piij}
\end{gathered}
\end{equation}
The average over any function $Q$ of $u$ in the plasma frame, or a function $ Q $ of $ u' $ in the pulsar frame is written as
\be
    n\langle Q\rangle
        = \int{\rm d}u\,Q\,g(u),
        \quad{\rm and}\quad
    n'\langle Q\rangle'
        = \int{\rm d}u'\,Q\,g'(u').
    \label{average}
\ee 
The Lorentz transformation of equation~\eqref{Piij} gives \citep{Me73}
\begin{equation}
\begin{split}
    \Pi_{1'1'} 
        & = \Pi_{11} = \Pi_{2'2'} = \Pi_{22} 
        = \frac{\omega'^2_{\rm p0}}{\Omega_{\rm e}^2}\frac{\omega'^2}{{z'}^2}\left\langle\gamma'(z'-\beta')^2\right\rangle',\\
    \Pi_{1'3'} 
        & = \Pi_{3'1'} = \gamma_{\rm s}\left[\Pi_{13}\frac{z+\beta_{\rm s}}{z}+\Pi_{11}\frac{\beta_{\rm s}\tan\theta}{z}\right]\\
        & = \gamma_{\rm s}\frac{\omega_{\rm p}^2}{\Omega_e^2}\frac{\omega^2\tan\theta}{z^2}\left\langle\gamma(z-\beta)(\beta_{\rm s} + \beta)\right\rangle
        = \frac{\omega'^2_{\rm p0}}{\Omega_e^2}\frac{\omega'^2\tan\theta'}{{z'}^2}\left\langle\gamma'\beta'(z' - \beta')\right\rangle',\\\ms
    \Pi_{3'3'} 
        & = \gamma_{\rm s}^2\left[\Pi_{33}\left(\frac{z+\beta_{\rm s}}{z}\right)^2+2\Pi_{13}\frac{z+\beta_{\rm s}}{z}\frac{\beta_{\rm s}\tan\theta}{z}+\Pi_{11}\left(\frac{\beta_{\rm s}\tan\theta}{z}\right)^2\right]\\
        & = -\omega_{\rm p}^2\gamma_{\rm s}^2(z + \beta_{\rm s})^2\left\langle\frac{1}{\gamma^3(z-\beta)^2}\right\rangle + \frac{\omega_{\rm p}^2}{\Omega_e^2}\gamma_{\rm s}^2\frac{\omega^2\tan^2\theta}{z^2}\left\langle\gamma(\beta_{\rm s} + \beta)^2\right\rangle\\
        & = -\omega'^2_{\rm p0} {z'}^2\left\langle\frac{1}{\gamma'^3(z'-\beta')^2}\right\rangle' + \frac{\omega'^2_{\rm p0}}{\Omega_e^2}\frac{\omega'^2\tan^2\theta'}{{z'}^2}\left\langle\gamma'{\beta'}^2\right\rangle'.
    \label{LT4}
\end{split}
\end{equation}
We provide explanation for the derivation of the above in Section~\ref{sec:LTexp}. The frequency $\Omega_{\rm e}$ is unchanged by the Lorentz transformation, and $ \omega'_{\rm p0} $ is the plasma frequency corresponding to the Goldreich-Julian number density,
\begin{equation}
    \omega'_{\rm p0}
        \approx 3.2\times10^8\, \left[\left(\frac{\kappa}{10^5}\right) \left(\frac{{\dot P}/P^3}{10^{-15}~{\rm s}^{-3}}\right)^{1/2}\left(\frac{r/r_L}{0.1}\right)^{-3}\left(\frac{1~{\rm s}}{P}\right)^2\right]^{1/2}{\rm rad\,s}^{-1},
\end{equation}
where $ P $ is the pulsar period, $ \dot{P} $ is its period derivative, $ \kappa $ is the muliplicity, $ r_L $ is the light cylinder radius, and $ r $ is radial distance. The plasma frequency in the plasma rest frame, $ \omega_{\rm p} $, is related to $ \omega'_{\rm p0} $ through the relation $n'=\gamma_{\rm s}n$ between the number densities in the two frames, implying $\omega'^2_{\rm p}=\gamma_{\rm s}\omega_{\rm p}^2$. To make the dependence on $ \gamma_{\rm s} $ explicit, we make the replacement $ \omega_{\rm p}^2 = \omega'^2_{\rm p0}/\gamma_{\rm s} $ with $ \omega'_{\rm p0} $ independent of $ \gamma_{\rm s} $.

\subsection{Evaluation of the response tensor in the pulsar frame}
\label{sec:LTexp}

The average of a quantity $ Q $ in the plasma rest frame, $ \langle Q\rangle $, may be related to its average value in the pulsar frame, $ \langle Q\rangle' $, using~\eqref{average} as
\be 
    n'\left\langle\frac{Q}{\gamma'}\right\rangle'
        = n\left\langle\frac{Q}{\gamma}\right\rangle,
    \label{Qav}
\ee 
with the number density in ${\cal K}'$ related to that in ${\cal K}$ by $n'=\gamma_{\rm s}n$. The relation (\ref{Qav}) is equivalent to the relation
\be 
   \langle{\tilde Q}\rangle'
        = \langle{\tilde Q}(1+\beta\beta_{\rm s})\rangle,\quad
    \langle{\tilde Q}\rangle
        = \gamma_{\rm s}^2\langle{\tilde Q}(1-\beta'\beta_{\rm s})\rangle'
    \label{Qavp}
\ee 
for any quantity ${\tilde Q}$. Examples include
\be
    \langle\beta'\rangle' 
        = \beta_{\rm s},\quad
   \langle\gamma'\rangle'
        = \gamma_{\rm s}\left[\langle\gamma\rangle +\langle\gamma\beta^2\rangle\beta_{\rm s}^2\right],\quad
    \langle1/\gamma'\rangle'
        = \langle1/\gamma\rangle/\gamma_{\rm s}.
\label{Qavps}
\ee

In transforming the particle distribution between frames we assume that $\rho$ and $n$ are parameters, defined by their physical interpretation in ${\cal K}$.\footnote{Some authors choose to write the exponent in the streaming distribution (\ref{Juttnerp}) as $-\rho'\gamma'(1-\beta'\beta_{\rm s})$ with $\rho'=\gamma_{\rm s}\rho$ implying that the temperature in ${\cal K}'$ is $T'=T/\rho_{\rm s}$ \citep{LSS10}.} With $n$ and $\rho$ regarded as parameters in the Lorentz transformation, $g'(u')=g(u)$ implies that the ratio of $du'g'(u')$ to $du\,g(u)$ is $du'/du=\gamma'/\gamma$.

In deriving~\eqref{Piij}, \eqref{LT4}, \eqref{eq:Lambdaij_final} and related results we use~\eqref{Qav} with the following identities
\begin{equation}\label{LT5}
\begin{gathered}
    \frac{\gamma_{\rm s}(z+\beta_{\rm s})}{z}
        = \frac{z'}{\gamma_{\rm s}(z'-\beta_{\rm s})},\quad
    \gamma_{\rm s}(1+z\beta_{\rm s}) 
        = \frac{1}{\gamma_{\rm s}(1-z'\beta_{\rm s})},\quad
    \frac{\beta_{\rm s}\tan\theta}{z} 
        = \frac{\beta_{\rm s}\tan\theta'}{\gamma_{\rm s}(z'-\beta_{\rm s})},\\\ms
    \frac{\omega'^2\gamma'^2\beta'^2\tan^2\theta'}{z'^2} 
        = \gamma_{\rm s}^2\frac{\omega^2\gamma^2(\beta+\beta_{\rm s})^2\tan^2\theta}{z^2},\quad
    \frac{\omega'^2\gamma'^2(z'-\beta')^2}{z'^2} 
        = \frac{\omega^2\gamma^2(z-\beta)^2}{z^2},\\\ms
    \frac{\omega'^2}{\omega^2} 
        = \left(\frac{z'}{\gamma_{\rm s}(z'-\beta_{\rm s})}\right)^2,\quad
    \frac{\omega'^2\gamma'^2(z'-\beta')\beta'\tan\theta'}{z'^2} 
        = \gamma_{\rm s}\frac{\omega^2\gamma^2(z-\beta)(\beta+\beta_{\rm s})\tan\theta}{z^2}.
\end{gathered}
\end{equation}

In particular using $ {\gamma'}^2(z' - \beta')^2 = \gamma^2(z - \beta)^2/\gamma_{\rm s}^2(1 + z\beta_{\rm s})^2 $ we may relate the RPDF $W'(z') $ in the pulsar frame to that in plasma rest frame, $ W(z) $, as
\be 
    W'(z') 
        = \left\langle\frac{1}{\gamma'^3(z'-\beta')^2}\right\rangle'
        = \gamma_{\rm s}(1+z\beta_{\rm s})^2\left\langle\frac{1}{\gamma^3(z-\beta)^2}\right\rangle 
        = \gamma_{\rm s}(1+z\beta_{\rm s})^2W(z).
\label{W'z'}
\ee 
It follows that the RPDFs in the two frames are related by
\be
    W'(z') 
        = \frac{1}{\gamma_{\rm s}^3(1-z'\beta_{\rm s})^2}\,W\left(\frac{z'-\beta_{\rm s}}{1-z'\beta_{\rm s}}\right),\quad
    W(z) 
        = \frac{1}{\gamma_{\rm s}(1+z\beta_{\rm s})^2}\,W'\left(\frac{z+\beta_{\rm s}}{1+z\beta_{\rm s}}\right).
\label{W'z'a}
\ee
We also note the relation
\be
    \frac{z'^2W'(z')}{\omega'^2}
        = \frac{1}{\gamma_{\rm s}}\frac{z^2W(z)}{\omega^2}.
\label{W'z'b}
\ee

\section{Wave modes in the pulsar frame}
\label{sect:dielectric}

Wave dispersion in the pulsar frame, ${\cal K}'$, may be treated in several different ways. One approach is to perform the calculations in ${\cal K}$, where the distribution function is $g(u)$, and Lorentz transform the wave properties to ${\cal K}'$. A second approach is to perform the calculations entirely in ${\cal K}'$ where the distribution function is $g'(u')$. A third approach is to Lorentz transform the dielectric tensor, evaluated in ${\cal K}$, to ${\cal K}'$. These three approaches are formally equivalent, and comparison of them facilitates identifying the relation between the RPDFs in the two frames.

\subsection{Transformed dispersion equation}

Treating wave dispersion in ${\cal K}'$ involves evaluating the tensor $K'_{i'j'}(\omega',{\bi k}')$, constructing the tensor $\Lambda'_{i'j'}(\omega',{\bi k}')$, where
\begin{equation}
\Lambda'_{i'j'}(\omega',{\bi k}') = \frac{c^2(k_{i'}k_{j'}-|{\bi k}'|^2\delta_{i'j'})}{\omega'^2} + K'_{i'j'}(\omega',{\bi k}'),
\end{equation}
and setting its determinant to zero to obtain the dispersion equation, which is 
\begin{equation}\label{eq:disp0}
\Lambda'_{2'2'}(\Lambda'_{1'1'}\Lambda'_{3'3'}-\Lambda'^2_{1'3'})=0.
\end{equation}
The explicit expressions for the nonzero components of $ \Lambda'_{i'j'} $ are
\begin{equation}\label{eq:Lambdaij_final}
\begin{split}
    \Lambda'_{1'1'} 
        & = 1 - \frac{1}{{z'}^2} + \frac{\omega'^2_{\rm p0}}{\Omega_e^2}\frac{1}{{z'}^2}\left\langle\gamma'(z' - \beta')^2\right\rangle',\\
    \Lambda'_{2'2'} 
        & = 1 - \frac{1}{{z'}^2\cos^2\theta'} + \frac{\omega'^2_{\rm p0}}{\Omega_e^2}\frac{1}{{z'}^2}\left\langle\gamma'(z' - \beta')^2\right\rangle'
        = \Lambda'_{1'1'} - \frac{\tan^2\theta'}{{z'}^2},\\
    \Lambda'_{1'3'} 
        & = \frac{\tan\theta'}{{z'}^2}\left[1 + \frac{\omega'^2_{\rm p0}}{\Omega_e^2}\left\langle\gamma'\beta'(z' - \beta')\right\rangle'\right],\\
    \Lambda'_{3'3'} 
        & = 1 -\frac{\omega'^2_{\rm p0}}{\omega'^2} {z'}^2\left\langle\frac{1}{\gamma'^3(z'-\beta')^2}\right\rangle' - \frac{\tan^2\theta'}{{z'}^2}\left[1 - \frac{\omega'^2_{\rm p0}}{\Omega_e^2}\left\langle\gamma'{\beta'}^2\right\rangle'\right].
\end{split}
\end{equation}
These are of the same form as those in the plasma rest frame (Paper~1). A detailed comparison shows
\begin{equation}
\omega'^2\Lambda'_{2'2'}=\omega^2\Lambda_{22}
\qquad
\omega'^4(\Lambda'_{1'1'}\Lambda'_{3'3'}-\Lambda'^2_{1'3'})=\omega^4(\Lambda_{11}\Lambda_{33}-\Lambda_{13}^2).
\end{equation}
It follows that the dispersion equations in the two frames are proportional to each other. 

A wave mode is a specific solution of the dispersion equation, and the transformation of the dispersion equation implies that there is a one-to-one correspondence between the solutions for wave modes in the two frames. One is free to identify the wave modes in ${\cal K}$ and use this one-to-one correspondence to identify the modes in ${\cal K}'$. However, the identification of specific modes is not uniquely defined, and this is only one possible prescription for identifying the modes in ${\cal K}'$. For example, waves in an arbitrary wave mode labeled $M$ with $\omega=\omega_M(z)>0$ in ${\cal K}$ may transform into $\omega'<0$ for some range of $z'$ in ${\cal K}'$ and be interpreted as defining a separate mode in ${\cal K}'$; we illustrate such a case below for the L~mode.

\subsection{Transformed dispersion relations}

We illustrate here the transformation of the dispersion relations from the rest frame ${\cal K}$ to the pulsar frame ${\cal K}'$. We first consider the X~mode, and the A~mode which is degenerate with the X~mode for parallel propagation; we then consider the L~mode and its generalization to the oblique propagation. 

The dispersion relations for parallel propagation ($ \theta = 0 $) in ${\cal K}$ are $z= \pm z_{\rm A}$ for the forward- and backward-propagating A~and X~mode waves, and $\omega = \pm \omega_{\rm L}(z)$ for the forward- and backward-propagating L~mode waves (Paper~1). Similarly, for parallel propagation in ${\cal K}'$ it is convenient to solve the dispersion equation for $z'$ when considering the A~and X~modes, and for $\omega'$ when considering the L~mode. 

The dispersion equation~\eqref{eq:disp0} in ${\cal K}'$ factorizes into  $ \Lambda'_{2'2'} = 0 $ for the X~mode and $ \Lambda'_{1'1'}\Lambda'_{3'3'}-\Lambda'^2_{1'3'} = 0 $ for the O~and Alfv\'en modes for oblique propagating waves. For parallel propagation the dispersion equation is $ \Lambda'_{1'1'} =\Lambda'_{2'2'} = 0 $ for the A~mode and the X~mode, and $ \Lambda'_{3'3'} = 0 $ for the L~mode. 

For oblique propagation, $ \theta' \neq 0 $, the X~mode is given by the dispersion equation $ \Lambda'_{2'2'} = 0 $ which is a quadratic equation in $z'$ with solutions $z'=z'_\pm$ where
\begin{equation}\label{eq:Xmode}
\begin{split}
    z'_\pm
        & = \frac{\omega'^2_{\rm p0}\langle\gamma'\beta'\rangle'/\Omega_e^2 \pm \Delta'^{1/2}}{1 + \omega'^2_{\rm p0}\langle\gamma'\rangle'/\Omega_e^2},
        \\
    \Delta' 
        & = \left(1 + \omega'^2_{\rm p0}\langle\gamma'\rangle'/\Omega_e^2\right)\left(\sec^2\theta'- \omega'^2_{\rm p0}\langle\gamma'{\beta'}^2\rangle'/\Omega_e^2\right) + \left(\omega'^2_{\rm p0}\langle\gamma'\beta'\rangle'/\Omega_e^2\right)^2.
\end{split}
\end{equation}
For $\gamma_{\rm s}\gg1$ the averages in (\ref{eq:Xmode}) are approximately equal, where we use~\eqref{W'z'b} to find $ \langle\gamma'\beta'^2\rangle' \approx \langle\gamma'\beta'\rangle' \approx \langle\gamma'\rangle' \approx 2\gamma_{\rm s}\langle\gamma\rangle $. It is convenient to write $a=2\gamma_{\rm s}\langle\gamma\rangle\omega'^2_{\rm p0}/\Omega_e^2$, such that one has $\Delta' \approx (1+a)\sec^2\theta'-a$, and then $\sec^2\theta'\ge1$ implies $\Delta'>0$ and that the solutions $ z'_\pm $ are always real. The solution $ z'_+ $ is always positive, and $ z'_- $ is positive for $\sec^2\theta'<a$ and negative for $\sec^2\theta'>a$. We note that
\begin{equation}\label{eq:xi}
   \frac{1}{\omega'^2_{\rm p0}/\Omega_e^2}
        \approx 2.6\times10^5 \left(\frac{\dot{P}/P^3}{10^{-15}~{\rm s}^{-3}}\right)^{1/2}\left(\frac{r/r_L}{0.1}\right)^{-3}\left(\frac{10^5}{\kappa}\right),
\end{equation}
and for $ \gamma_{\rm s} = 10^{2}{\rm -}10^{3} $ and $ \langle\gamma\rangle = 1{\rm -}10^2 $ the condition $ a > \sec^2\theta' $ or $ \gamma_{\rm s}\langle\gamma\rangle \gtrsim \sec^2\theta'/(2\omega'^2_{\rm p0}/\Omega_e^2) $ cannot be satisfied. Therefore the X~mode always exists with $ z'_+ $ corresponding to the forward-propagating mode, and $ z'_- $ corresponding to the backward-propagating mode except for $ \gamma_{\rm s}\langle\gamma\rangle \gtrsim \sec^2\theta'/(2\omega'^2_{\rm p0}/\Omega_e^2) $. The solution for the X~mode in the plasma frame ${\cal K}$ is obtained by omitting the primes in~\eqref{eq:Xmode}, making the replacement $ \omega'^2_{\rm p0} \to \omega_{\rm p}^2 $ and noting that  $ g(u) $ is an even function of $ \beta $ in ${\cal K}$ implying $ \langle Q\rangle = 0 $ if $ Q $ is an odd function of $ \beta $. It follows that in ${\cal K}$ one has $ z_+ = -z_- $, and the solutions are interpreted as forward-propagating and backward-propagating X~mode waves, respectively. For parallel-propagating X~mode and for the A~mode these solutions correspond to $z_\pm=\pm z_A$ in $ \mathcal{K} $.

For the purpose of comparison of the X~mode in the two frames it is convenient to write $ \sec^2\theta' $ in terms of $ \tan^2\theta $ using~\eqref{LT2}. Then in~\eqref{eq:Xmode} one makes the replacements $ \sec^2\theta'  \to 1 + \gamma_{\rm s}^2\tan^2\theta$,
\begin{equation}\label{eq:Lambda11b}
    \frac{\omega'^2_{\rm p0}}{\Omega_e^2}\langle\gamma'\rangle' 
        \to \frac{\omega'^2_{\rm p0}}{\Omega_e^2}\langle\gamma'\rangle' - \gamma_{\rm s}^2\beta_{\rm s}^2\tan^2\theta,\quad
    \frac{\omega'^2_{\rm p0}}{\Omega_e^2}\langle\gamma'\beta'\rangle'
        \to \frac{\omega'^2_{\rm p0}}{\Omega_e^2}\langle\gamma'\beta'\rangle' - \gamma_{\rm s}^2\beta_{\rm s}\tan^2\theta.
\end{equation}

The dispersion equation for the A~mode for parallel propagation in ${\cal K}'$ is $ \Lambda'_{1'1'} = 0 $. The solution of $ \Lambda'_{1'1'} = 0 $ is obtained by setting $ \cos^2\theta' = 1 $ in the solution of $ \Lambda'_{2'2'}=0 $, with the X~mode and A~mode degenerate for parallel propagation, as in ${\cal K}$. Specifically, the A~mode (and also the parallel X~mode) is given by
\begin{equation}\label{eq:Amode}
    \left.z'_1 = z_+\right|_{\theta' = 0},\quad
    \left.z'_2 = z_-\right|_{\theta' = 0}.
\end{equation}
The solutions obtained from $ z'_\pm $ remain forward/backward-propagating except for $a>1$, that is for $ \gamma_{\rm s}\langle\gamma\rangle \gtrsim 1/(2\omega'^2_{\rm p0}/\Omega_e^2) $, in which case the backward-propagating mode becomes forward-propagating. This may be interpreted in terms of $z=-z_A$ transforming into $z'=(-z_A+\beta_{\rm s})/(1-z_A\beta_{\rm s})>0$.

The dispersion equation for the transformed L~mode (for parallel propagation) is $ \Lambda'_{3'3'} = 0 $ which, using~\eqref{eq:Lambdaij_final}, may be written as
\begin{equation}\label{eq:Lmode}
    \omega'^2 = \omega'^2_L(z'),\quad
    \omega'^2_L(z') = \omega'^2_{\rm p0}z'^2W'(z').
\end{equation}
The solution in ${\cal K}$, $\omega^2=\omega_L^2(z)$, includes both positive and negative frequencies which may be chosen to be $\omega=\pm\omega_L(\pm z)$, with $\omega_L^2(-z)=\omega_L^2(z)$. The negative-frequency solution for $z>0$ may be interpreted as a positive-frequency solution for $z<0$, that is, for a backward-propagating wave. The dispersion equation in ${\cal K}$ then has symmetric peaks at $z=\pm z_{\rm m}$ and zeroes at $z=\pm z_0$, with a region $-z_0<z<z_0$ of evanescence where there are no propagating waves. The Lorentz-transformed dispersion equation in ${\cal K}'$ may be interpreted as a shifted, distorted version of this dispersion equation in ${\cal K}$: the peak at $z=z_{\rm m}$ is enhanced and shifted nearer to $z=1$, the region of evanescence centered on $z=0$ is shifted to large $z$, and the peak at $z=-z_{\rm m}$ is diminished and also shifted to larger $z$. If $z=-z_0$ transforms to $z'>0$ the negative-$\omega$ solution becomes a positive-$\omega'$ solution of the L~mode, which includes the diminished peak if $z=-z_{\rm m}$ transforms to $z'>0$. One may either interpret this in terms of two branches of L~mode, both with $\omega'>0$ and $z'>0$. Alternatively one may interpret these as two separate L~modes in ${\cal K}'$.

\subsection{Oblique modes in ${\cal K}'$}
\label{sec:oblique}

For oblique propagation in ${\cal K}'$, the solution of $ \Lambda'_{1'1'}\Lambda'_{3'3'}-\Lambda'^2_{1'3'} = 0 $ is given by
\begin{equation}\label{eq:disp1}
    \omega'^2 = \frac{(z' - z'_1)(z' - z'_2)\,{\omega'}_L^2(z')}{(z' - z'_1)(z' - z'_2) - b'\tan^2\theta'},
\end{equation}
where $ z'_1 $ and $ z'_2 $ are roots of $ \Lambda'_{1'1'} = 0 $ given by~\eqref{eq:Amode} and $ \omega'^2_L(z') $ is given by~\eqref{eq:Lmode}, and
\begin{equation}
\begin{split}
    b' 
        & = 1 + \frac{\omega'^2_{\rm p0}}{\Omega_e^2}\left\langle1/\gamma'\right\rangle' + \frac{{\omega'}_{\rm p0}^4}{\Omega_e^4}\left[\left\langle\gamma'\beta'\right\rangle'^2 - \left\langle\gamma'\right\rangle'\left\langle\gamma'\beta'^2\right\rangle'\right]\\
        & = 1 + \frac{\omega'^2_{\rm p0}}{\Omega_e^2}\frac{\left\langle1/\gamma\right\rangle}{\gamma_{\rm s}} - \frac{{\omega'}_{\rm p0}^4}{\Omega_e^4}\left[\left\langle\gamma\right\rangle^2 - \left\langle\gamma\right\rangle\left\langle1/\gamma\right\rangle\right],
\end{split}
\end{equation}
with $ b' \approx 1 $ for $ \langle\gamma\rangle \ll 1/(\omega'^2_{\rm p0}/\Omega_e^2) $, which is well satisfied for pulsar parameters. For comparison in the plasma frame we may drop the primes in~\eqref{eq:disp0} and make the replacement $ b' \to b = 1 - \omega_{\rm p}^2 \langle\gamma\beta^2\rangle/\Omega_e^2 \approx 1 $, and $ |z'_{1,2}| \to z_A $ with $ z_A^2 = (1 - \omega_{\rm p}^2\langle\gamma\beta^2\rangle/\Omega_e^2)/(1 + \omega_{\rm p}^2\langle\gamma\rangle/\Omega_e^2)$.

\begin{figure}
\begin{center}
\psfragfig[width=1.0\columnwidth]{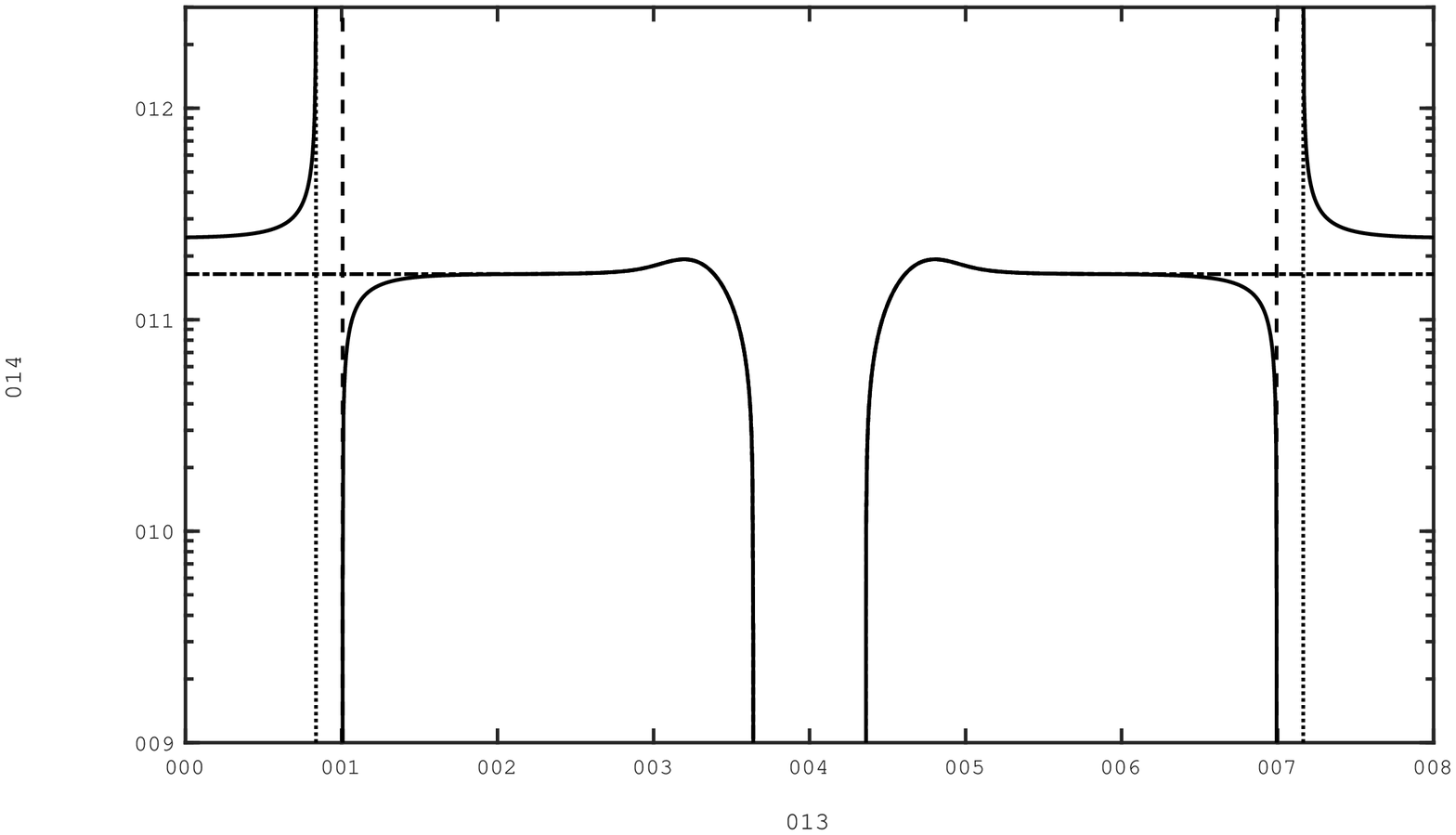}\\\vspace{3mm}
\psfragfig[width=1.0\columnwidth]{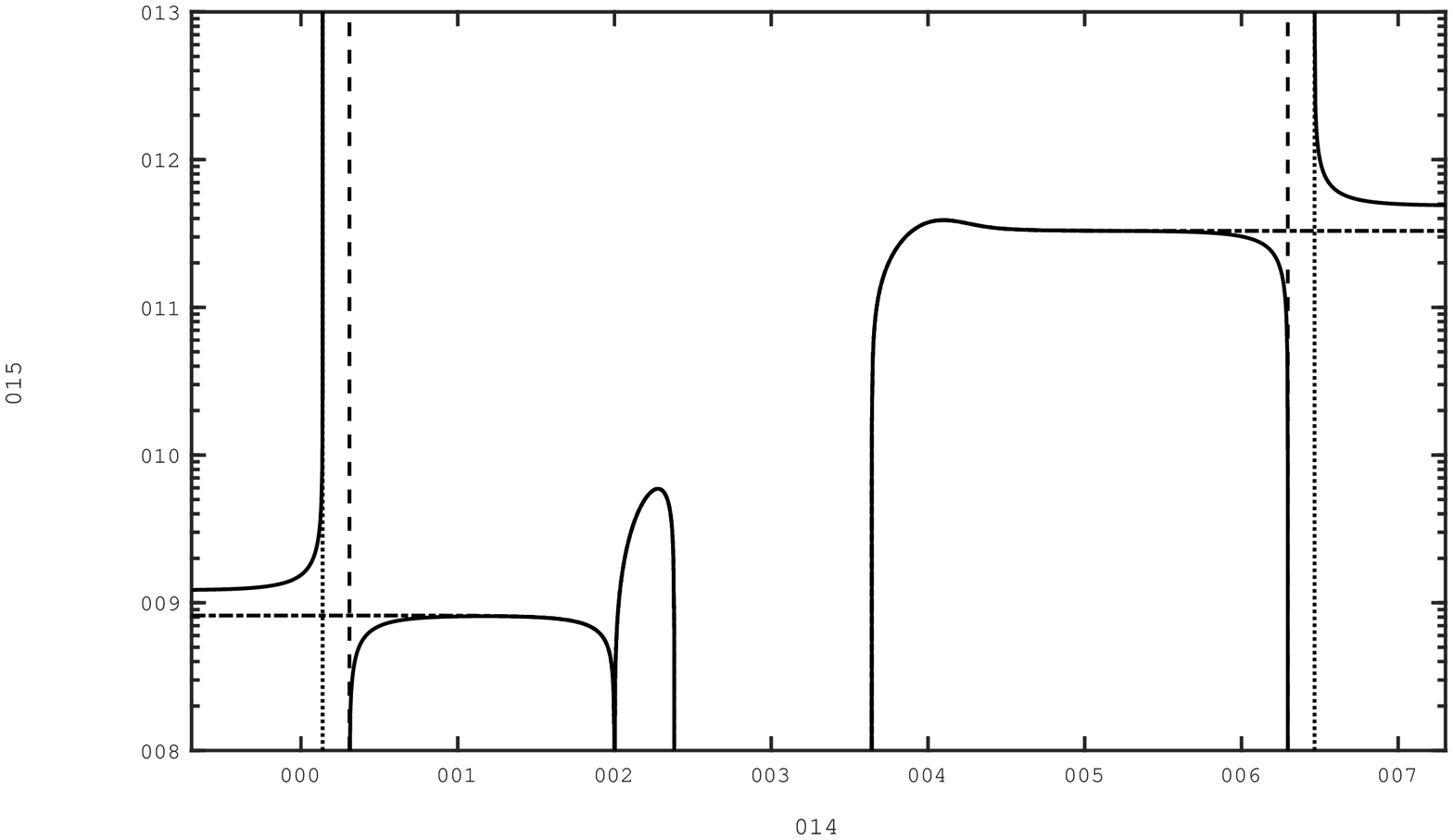}
 \caption{TOP (plasma frame): plot of $ \omega_L(z)/\omega'_{\rm p0} $ (dash-dotted), $ \omega/\omega'_{\rm p0} $ (solid) as given by counterpart of~\eqref{eq:disp0} in Paper~1, $ z = \pm z_A/\cos\theta $ (dotted) and $ z = \pm z_A $ (dashed). BOTTOM (pulsar frame): plot of $ \omega'_L(z)/\omega'_{\rm p0} $ (dash-dotted), $ \omega'/\omega'_{\rm p0} $ (solid) as given by~\eqref{eq:disp0}, $ z' = z'_\pm $ (dotted) and $ z' = z'_{1,2} $ (dashed). For all plots we use $ \rho = 1 $, $ \gamma_{\rm s} = 10 $ and $ \theta = 7.5\times10^{-4} $. In both figures, forward and backward propagating waves are to the right and left of zero, respectively. Only subluminal waves are included in these figures.}
  \label{fig:ModesPlasmaPulsar} 
 \end{center}
\end{figure}

With $ {\omega'}_L^2(z') \geq 0 $, the numerator of~\eqref{eq:disp0} is greater than zero for $ z'> z'_1 $ and for $ z' < z'_2 $, and the denominator is greater than zero for $ z' > \bar{z}'_+ $ and for $ z' < \bar{z}'_- $, with
\begin{equation}
    \bar{z}'_\pm 
        = \frac{1}{2}(z'_1 + z'_2) \pm \frac{1}{2}\left[(z'_1 - z'_2)^2 + 4b'\tan^2\theta'\right]^{1/2} 
        \approx
        \begin{pmatrix}
        z'_1\\
        z'_2
        \end{pmatrix}
        \pm \frac{b'\tan^2\theta'}{(z'_1 - z'_2)},
\end{equation}
where the approximation applies for $ b'\tan^2\theta' \ll (z'_1 - z'_2)^2 $. As discussed in Paper~1, the solutions $ \omega^2={\omega}_L^2(z) $ are real except for $ -z_0 <z<z_0$, with $ z_0 \approx 1 - 0.14\rho^2 $ for $ \rho \ll 1 $ and $ z_0 \approx 0.9 $ for $\rho=1$, and they are weakly damped for $z\lesssim -z_{\rm m}$, $z\gtrsim z_{\rm m}$, where $z=\pm z_{\rm m}$, with $z_{\rm m}\approx1 - 0.013\rho^2$ for $ \rho \ll 1 $ and $ z_{\rm m} \approx 1-0.0124 $ for $\rho=1$, corresponds to symmetric maxima in the RPDF. In the pulsar frame one has $ {\omega'}_L^2(z') > 0 $ for $ z' < z'_{0-} $ and $ z' > z'_{0+} $ where 
\begin{equation}\label{eq:z'0pm}
    z'_{0\pm} = \frac{\beta_{\rm s} \pm z_0}{1 \pm \beta_{\rm s} z_0},
    \qquad
    z'_{\rm m\pm} = \frac{\beta_{\rm s} \pm z_{\rm m}}{1 \pm \beta_{\rm s} z_{\rm m}}.
\end{equation}
It follows that for $ {\omega'}_L^2(z') > 0 $ a solution of~\eqref{eq:disp0} exists over domain $ D'_1 $ and for $ {\omega'}_L^2(z') < 0 $ a solution exists over domain $ D'_2 $ where
\begin{align}
    D'_1 
        & = \left\{(-\infty, \bar{z}'_-) \cup (z'_2, z'_1) \cup (\bar{z}'_+, \infty)\right\}\cap\left\{(-\infty, z'_{0-}) \cup (z'_{0+}, \infty)\right\},\\
    D'_2
        & = \left\{(\bar{z}'_-, z'_2) \cup (z'_1, \bar{z}'_+)\right\} \cap (z'_{0-}, z'_{0+}).
\end{align}
As discussed in Paper~1 the solution of the counterpart of~\eqref{eq:disp0} in the plasma frame is heavily damped except to $ |z|\gtrsim z_{\rm m} $. This is also the case in the pulsar frame for solutions over domain $ D'_2 $ and as such we do not discuss it here. Explicitly, for domain $ D'_1 $ we have L~mode over $ z' \geq z'_{0+} $ and $ z' \leq z'_{0-} $, A~mode at $ z' = z'_1 $ and at $ z' = z'_2 $, X~mode at $ z' = z'_+ $ and at $ z' = z'_- $, Alfv\'en mode over $ z'_{0+} \leq z' < z'_1 $ and $ z'_2 < z' \leq z'_{0-} $, and O~mode over $ z' > \bar{z}'_+ $ and $ z' < \bar{z}'_- $.

From~\eqref{eq:z'0pm} we see that $ z'_{0+} > 0 $ always and that $ z'_{0-} < 0 $ except for $ \beta_{\rm s} \geq z_0 $ which corresponds to $ \gamma_{\rm s} \geq \gamma_0 $ where $ \gamma_0 \approx 1.9\langle\gamma\rangle $ for $ \rho \ll 1 $ and $ \gamma_0 \approx 2.3 $ for $ \rho = 1 $. This implies that for $ \gamma_{\rm s} \geq \gamma_0 $ part of the backward-propagating wave in the plasma frame becomes forward propagating in the pulsar frame. 

Figure~\ref{fig:ModesPlasmaPulsar} shows plots of the plasma dispersion in the plasma rest frame (TOP) and the pulsar frame (BOTTOM) for $ \rho = 1 $, $ \gamma_{\rm s} = 10 $ and $ \theta = 7.5\times10^{-4} $. The top panel (plasma frame) shows plot of the L~mode $ \omega_L(z)/\omega'_{\rm p0} $ (dash-dotted), O~mode and Alfv\'en~mode $ \omega/\omega'_{\rm p0} $ (solid) as given by counterpart of~\eqref{eq:disp0} in Paper~1, X~mode $ z = \pm z_A/\cos\theta $ (dotted), and A~mode $ z = \pm z_A $ (dashed). The bottom panel (pulsar frame) shows the corresponding Lorentz transformed L~mode $ \omega'_L(z)/\omega'_{\rm p0} $ (dash-dotted), O~mode and Alfv\'en~mode $ \omega'/\omega'_{\rm p0} $ (solid) as given by~\eqref{eq:disp0}, X~mode $ z' = z'_\pm $ (dotted) given by~\eqref{eq:Xmode}, and A~mode $ z' = z'_{1,2} $ (dashed) obtained from the X~mode by setting $ \theta = 0 $. Corresponding curves in the two frames are related through a Lorentz transformation. 

In the pulsar frame the portion of the O~mode asymptotic to $ z = \bar{z}'_\pm $ may be approximated as
\begin{equation}
    z' \approx 
        \begin{pmatrix}
            z'_1\\
            z'_2
        \end{pmatrix}
        \pm \frac{b'\tan^2\theta'}{(z'_1 - z'_2)}\left(1 + \frac{\omega'^2_L(z')}{\omega'^2}\right),
\end{equation}
where we assume $ \omega'^2 \gg \omega'^2_L(z') $, $ (z'_1 - z'_2)^2 \gg b'\tan^2\theta' $, and treat $ \omega'^2_L(z') $ as a constant over this narrow range. 

\begin{figure}
\begin{center}
\psfragfig[width=1.0\columnwidth]{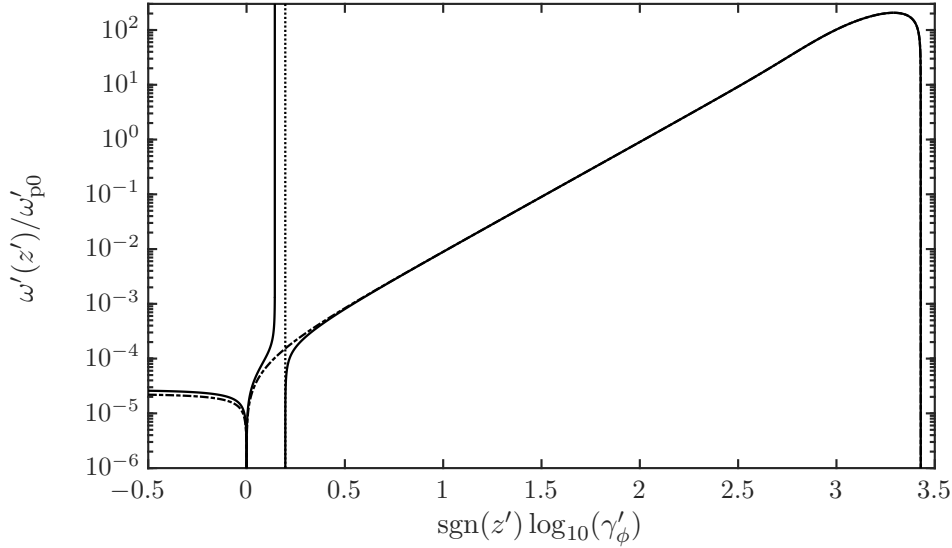}
 \caption{Plot of Lorentz boosted L~mode $ \omega'_L(z')/\omega'_{\rm p0} $ (dash-dotted), O and Alfv\'en modes (solid) and the X~mode (dotted) for $ \rho = 0.1 $, $ \gamma_{\rm s} = 10^5 $ and $ \theta = 5\times10^{-6} $. These modes are backward-propagating in the plasma rest frame.}
  \label{fig:ModesPlasmaPulsar2} 
 \end{center}
\end{figure}

The forward (backward) propagating O~mode is to the right (left) of the forward (backward) propagating X~mode.\footnote{Strictly to the right and left of $ \bar{z}'_+ $ and $ \bar{z}'_- $ respectively, however, $ z'_\pm \approx \bar{z}'_\pm $ for most parameter values.} The forward (backward) propagating Alfv\'en~mode is between the forward (backward) propagating A~mode and $ z'_{0+} $ ($z'_{0-}$). In Figure~\ref{fig:ModesPlasmaPulsar} the backward propagating L and Alfv\'en modes in the plasma rest frame are transformed partially to positive $ z' $ values, specifically to the loop immediately to the right of zero on the horizontal axis, with the top of the loop at $z'=z'_{m-}>0$, given by (\ref{eq:z'0pm}). As already remarked, one could consider this branch as an additional mode in ${\cal K}'$, but we prefer to regard it as a transformed part of the Alfv\'en or O~modes in ${\cal K}$. A more extreme case is shown in Figure~\ref{fig:ModesPlasmaPulsar2} where we show the L~mode $ \omega'_L(z')/\omega'_{\rm p0} $ (dash-dotted), O and Alfv\'en modes (solid) and the X~mode (dotted) for $ \rho = 0.1 $, $ \gamma_{\rm s} = 10^5 $ and $ \theta = 5\times10^{-6} $. We only show the portion that is backward-propagating in the plasma rest frame. For these parameter values the backward-propagating Alfv\'en  mode in the plasma frame is transformed completely to positive $ z' $ in the pulsar frame. Additionally the backward-propagating X~mode and O~mode are transformed to positive $ z' $. In this case one could say that there are six modes  with positive $ z' $ in ${\cal K}'$: Lorentz transformed O, Alfv\'en, X-modes with $z>0$ plus Lorentz transformed O, Alfv\'en, X~modes with $z<0$. 

We show how the single L~mode in ${\cal K}$ can seemingly split into two separate L~modes in ${\cal K}'$. The existence of two longitudinal modes in ${\cal K}'$ was identified by~\cite{BGI93}.  However, we argue in \S\ref{sect:fourth} that approximations made by~\cite{BGI93} in effectively evaluating the RPDF in ${\cal K}'$ are misleading concerning the specific forms for the two dispersion relations. The identification of an additional mode in ${\cal K}'$ is a matter of convention. We prefer the convention in which the (three) modes are defined in ${\cal K}$, with the L~mode appearing to split into two modes in ${\cal K}'$. 

\subsection{Approximate dispersion relation for the O~mode}

The approximate dispersion relation given by equation (5.8) of Paper~1 for the O~mode in ${\cal K}$ needs to be transformed to ${\cal K}'$ when treating ray tracing in ${\cal K}'$. The transformed approximate dispersion relation may be written
\be
    N'^2_{\rm O}
        \approx 1 - \frac{\omega_{\rm p}^2}{\langle\gamma\rangle\omega'^2}
        \left[\sin^2\theta(1 + 3\cos^2\theta)\right],
\label{drOmodep}
\ee 
with the expression in square brackets to be re-expressed in terms of primed variables. For $N^2_{\rm O}\approx1$, implying $N'^2_{\rm O}\approx1$, the transformation of the angles may be approximated by the familiar aberration formulae, for example, $\cos\theta=(\cos\theta' - \beta_{\rm s})/(1 - \beta_{\rm s}\cos\theta')$.

\section{Splitting of the modes in the pulsar frame}
\label{sect:fourth}

In~\ref{sec:oblique} we demonstrate how the X, Alfv\'en and O~modes in ${\cal K}$ each transform into two branches in ${\cal K}'$, such that each branch may be re-interpreted as an additional mode in ${\cal K}'$. The suggestion that there is a fourth longitudinal mode in the pulsar frame \citep{BGI93,LG-S06} is an example of this effect. However, this specific suggestion is complicated by the analysis leading to a fourth mode being based on an approximation, effectively to the RPDF in the pulsar frame ${\cal K}'$, that is misleading when applied to wave dispersion for $|1-z'|\ll1$. We comment on this approximation below.

\begin{figure}
\begin{center}
\psfragfig[width=1.0\columnwidth]{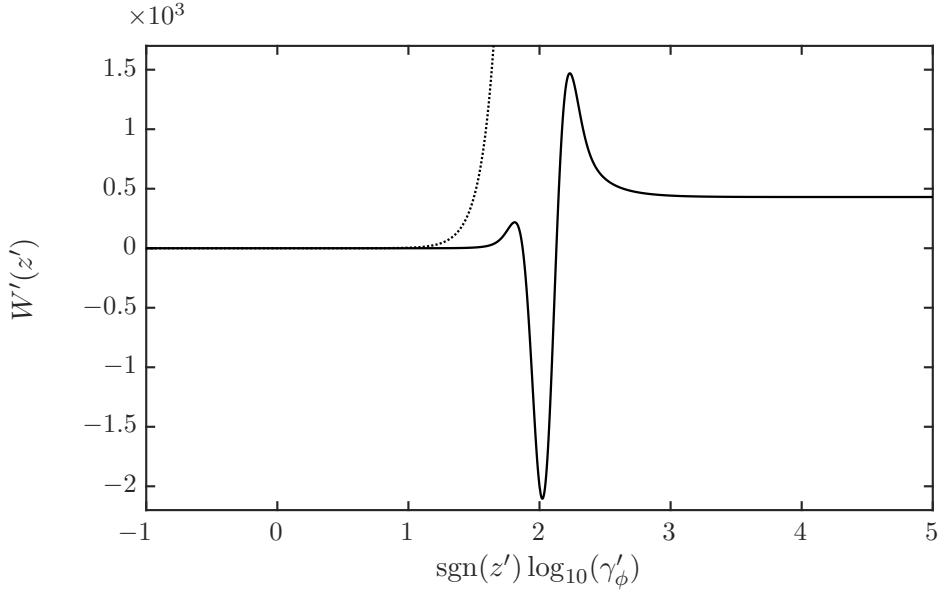}
 \caption{Plot of the exact RPDF $ W'(z') $ (solid) and the approximation~\eqref{eq:invapp} (dotted) in the subluminal region. We use $ \rho = 20 $ and $ \gamma_{\rm s} = 10^2 $.}
  \label{fig:W_approx} 
 \end{center}
\end{figure}

\subsection{Approximation to the RPDF in ${\cal K}'$}

In the pulsar frame the bulk velocity of the plasma is relativistic, $\gamma_{\rm s}\gg1$, implying that for $ \rho \gg 1 $ nearly all the particles have $\gamma'$ of order $ \gamma_{\rm s}$ and hence $1-\beta'$ of order $1/2\gamma_{\rm s}^2$. It seems plausible that the denominator $\beta'-z'$ in the RPDF (\ref{W'z'}) may be approximated by making the replacement $ \beta' \to 1 $ to lowest order in an expansion in $1/\gamma_{\rm s}^2$. This gives
\begin{equation}\label{eq:invapp}
    W'(z') = \left\langle\frac{1}{\gamma'^3(\beta' - z')^2}\right\rangle'
        \approx \frac{\langle1/\gamma'^3\rangle'}{(1 - z')^2}.
\end{equation}
This approximation is misleading in several ways. First, it is singular at $z'=1$, whereas the actual RPDF is always finite. Second, it is symmetric about $z'=1$, whereas the actual RPDF is asymmetric and sharply peaked, at $z'=z'_{\rm m\pm} $, with $1-z'_{\rm m+}\ll1$. These peaks in the RPDF play a dominant role in determining the properties of the L~mode (and the Alfv\'en and O~modes for small $\theta'$) for $1-z'\ll1$.

We show plots of the exact RPDF $ W'(z') $ (solid) and the approximation~\eqref{eq:invapp} (dotted) in the subluminal region in Figure~\ref{fig:W_approx}. We use $ \rho = 20 $ and $ \gamma_{\rm s} = 10^2 $. The portion of the graph to the left of $ \gamma'_\phi = \gamma_s = 10^2 $ corresponds to subluminal waves that are backward propagating in the plasma rest frame. From the symmetry of the approximation about $ z' = 1 $ we conclude that the approximation does not apply when $ |1 - z'| \lesssim 1/2\gamma_s $ in both the subluminal and superlumional regions. In particular, the approximation does not hold near the peaks of $ W'(z') $.

The plasma modes given by~\cite{BGI93} follow from the dispersion equation~\eqref{eq:disp0}, $ \Lambda'_{2'2'}(\Lambda'_{1'1'}\Lambda'_{3'3'} - \Lambda'^2_{1'3'}) = 0 $, by making the approximations
\begin{equation}\label{eq:param_approx}
    b' \approx 1,\quad
    z'^2_\pm \approx \sec^2\theta',\quad
    z'^2_{1,2} \approx 1,
\end{equation}
assuming $ \gamma_{\rm s}\langle\gamma\rangle \ll 1/(2\omega'^2_{\rm p0}/\Omega_e^2) $, and expressing $z'$ in terms of the refractive index $ N' = 1/z'\cos\theta' $. This gives~\cite{BGI93}
\begin{equation}\label{eq:Bdisp}
    \left(1 - N'^2\right)\left(1 - N'^2 - (1 - {N'}^2\cos^2\theta')\frac{\omega'^2_{\rm p0}}{\omega'^2}\bigg\langle\frac{1}{\gamma'^3(1 - \beta' N' \cos\theta')^2}\bigg\rangle'\right) = 0,
\end{equation}
The solution $N'^2=1$ corresponds to the X~mode in this approximation. We are concerned with the modes described by the other factor in (\ref{eq:Bdisp}), which implies the dispersion equation
\begin{equation}\label{eq:Betal}
    {N'}^2 = 1 - (1 - {N'}^2\cos^2\theta')\frac{\omega'^2_{\rm p0}}{\omega'^2}\left\langle\frac{1}{\gamma'^3(1 - \beta' N' \cos\theta')^2}\right\rangle'.
\end{equation}
The approximation (\ref{eq:invapp}) corresponds to setting $\beta' N' \cos\theta'=N' \cos\theta'$. \cite{BGI93} considered two cases: $ |1-N'| \ll 1/{\gamma'}^2 $ and $ 1 \gg |1-N'| \gg 1/{\gamma'}^2 $, and it is the latter case that corresponds to the approximation (\ref{eq:invapp}). We discuss only this case.

The dispersion equation~\eqref{eq:disp1} with approximations~\eqref{eq:param_approx} corresponds to~\eqref{eq:Betal} and using~\eqref{eq:invapp} may be written as
\begin{equation}\label{eq:Betal_v}
    \frac{\omega'^2}{\omega'^2_{\rm p0}}
        \approx \frac{z'^2\langle1/\gamma'^3\rangle'}{(z' - 1)(z' - \sec\theta')},
\end{equation}
which applies only for $ |1 - z'| \ll 1 $. The solutions of interest for $\theta'^2\ll1$ are
\begin{equation}\label{eq:Betal_3modes}
    N' = 1 + \frac{\theta'^2}{4} \pm \left[\frac{\omega'^2_{\rm p0}}{\omega'^2}\left\langle1/\gamma'^3\right\rangle' + \frac{\theta'^4}{16}\right]^{1/2},
\end{equation}
only one of which is subluminal. We comment on these two solutions in the discussion following Figure~\ref{fig:ModesPlasmaPulsar3}.

\begin{figure}
\begin{center}
\psfragfig[width=1.0\columnwidth]{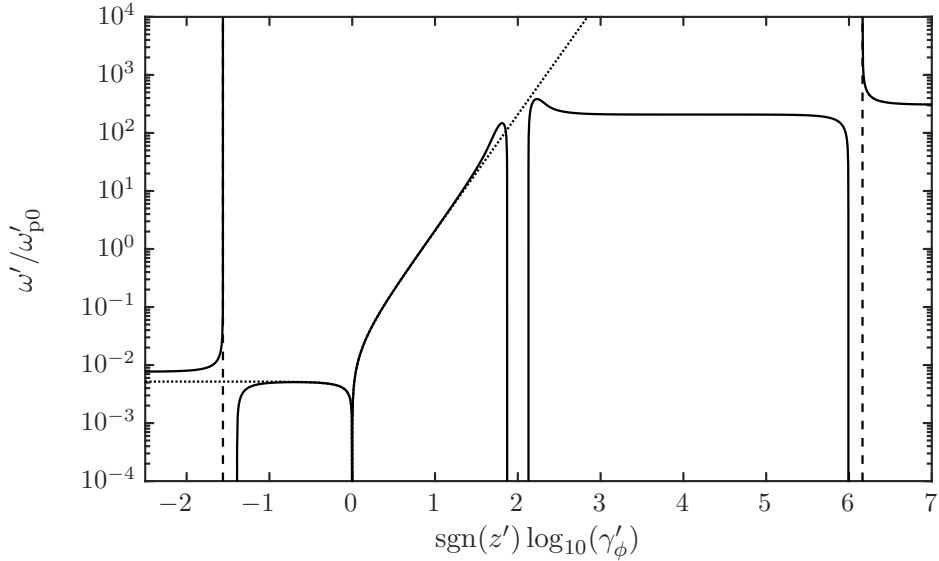}
 \caption{Plot of the exact X~mode (dashed), Alfv\'en mode (solid curves between the dashed lines) and O~mode (remaining solid curves). The dotted curve corresponds to the approximation~\eqref{eq:Betal_v} of~\cite{BGI93}. We use $ \rho = 20 $, $ \gamma_{\rm s} = 10^2 $ and $ \theta = 1.5\times10^{-4} $.}
  \label{fig:ModesPlasmaPulsar3} 
 \end{center}
\end{figure}

In Figure~\ref{fig:ModesPlasmaPulsar3} we show plots of the exact X~mode (dashed), Alfv\'en mode (solid curves between the dashed lines) and O~mode (remaining solid curves) in ${\cal K}'$ for $ \rho = 20 $, $ \gamma_{\rm s} = 10^2 $ and $ \theta = 1.5\times10^{-4} $. Waves to the left of $\log_{10}\gamma'_\phi=0$ are backward propagating in both ${\cal K}$ and ${\cal K}'$, waves in the range $0<\log_{10}\gamma'_\phi<2$ (i.e., $1<\gamma'_\phi<\gamma_{\rm s}$) are backward propagating in ${\cal K}$ and forward propagating in ${\cal K}'$, and waves to the right of $\log_{10}\gamma'_\phi=2$ are forward propagating in both ${\cal K}$ and ${\cal K}'$. In the subluminal range (we do not consider the superluminal range in Figure~\ref{fig:ModesPlasmaPulsar3}) it is apparent that all three modes split into separate branches, depending on pulsar parameters and whether they are forward or backward propagating in ${\cal K}$.

The dotted curve in Figure~\ref{fig:ModesPlasmaPulsar3} is one of the solutions derived by~\cite{BGI93} using the approximation to the RPDF that we criticize in \S\ref{sect:fourth}; specifically, the dotted curve corresponds to the subluminal Alfv\'en~mode in this approximation.  It is apparent that the dispersion relation corresponding to dotted curve is a good approximation to the actual dispersion relation only for $\gamma'_\phi\ll\gamma_{\rm s}$, but not for $\gamma'_\phi\gtrsim\gamma_{\rm s}^{1/2}$. In particular, it ignores the gap (around $\gamma'_\phi=\gamma_{\rm s}$) between the two branches of the Alfv\'en~mode, and it leads to a singularity at $\gamma'_\phi\to\infty$ ($z'\to1$). Another (superluminal) solution, cf.\ (\ref{eq:Betal_3modes}), is misleading for similar reasons, notably the singularity at $z'=1$ whereas the Alfv\'en~mode is well-behaved at $z=z'=1$. More specifically, the misleading approximation is that the RPDF may be approximated by a dependence $\propto1/(z'-1)^2$ near $z'=1$; this is an inappropriate approximation to the actual RPDF for $z'\approx1$.

\section{Discussion and conclusions}
\label{sect:conclusions}

In this paper we extend the discussion in Paper~1 of waves in the rest frame of a pulsar plasma to treat several problems that involve Lorentz transforming between frames. In \S\ref{sect:Lorentz} we discuss the transformation between the rest frame, ${\cal K}$, and the pulsar frame, ${\cal K}'$, in detail, emphasizing the transformation of the phase speed of the waves. 

In \S\ref{sect:beam} we apply a Lorentz transformation to an arbitrary 1D distribution (including a 1D J\"uttner distribution), $g(u)$, in the rest frame to derive the corresponding streaming distribution, $g'(u')=g(u)$ in ${\cal K}'$. We argue that relativistic streaming should be included in this way, that is, by applying a Lorentz transformation to a rest-frame distribution. A surprising implication is that such a Lorentz-transformed distribution is much broader (in ${\cal K}'$) than the original distribution (in ${\cal K}$). Specifically, a distribution confined to a range of $u$ of order $\langle\gamma\rangle\gg1$ in ${\cal K}$ is spread over a range of $u'$ of order $\gamma_{\rm s}\langle\gamma\rangle$ in ${\cal K}'$. In Paper~1 we emphasize the importance of including the relativistic spread in Lorentz factors, $\langle\gamma\rangle$, on the properties of wave dispersion, and in this paper we show that the effects of $\langle\gamma\rangle\gg1$ can be surprisingly large on the distribution function when the streaming is included. In particular, the transformed J\"uttner distribution, $g(u)\propto\exp(-\rho\gamma)$ transforms into the much broader distribution $\propto\exp[-\rho(\gamma_{\rm s}^2-\gamma^2)/2\gamma_{\rm s}\gamma]$. 

A conventional choice of a relativistically streaming distribution is a Gaussian distribution of the form (\ref{gup}), that is, $g(u)\propto\exp[-(u-u_\alpha)^2/2u_{\rm th}^2]$. We emphasize that this distribution is not the result of applying a Lorentz transformation to a Gaussian distribution $\propto\exp[-u^2/2u_{\rm th}^2]$ in the rest frame; the result of doing so is the distribution (\ref{gugsur}), which is much broader than the assumed relativistically streaming distribution. There is no obvious physical justification for such a relativistically streaming distribution, in preference to a distribution obtained by Lorentz-transforming a rest frame distribution. We adopt the view that a Lorentz-transformed J\"uttner distribution should be the preferred choice for a relativistically streaming distribution. 

In \S\ref{sect:separation} we discuss one effect of the broadness of a streaming distribution obtained by applying a Lorentz transformation to a J\"uttner distribution in the rest frame. The background and streaming distributions need to be well separated (in 4-speed) for the relativistic counterpart of the bump-in-tail instability to apply, and the broadness of the streaming distribution makes this separation condition much more restrictive than for say a relativistically streaming distribution $\propto\exp[-(u-u_\alpha)^2/2u_{\rm th}^2]$. An important implication is that the condition for waves to grow is much more restrictive than the conventional choice $\propto\exp[-(u-u_\alpha)^2/2u_{\rm th}^2]$ would imply.

In \S\ref{app:LT} we apply a Lorentz transformation to the dielectric tensor and derive the relation between the RPDF $W(z)$ in ${\cal K}$ and the transformed RPDF $W'(z')$ in ${\cal K}'$. 

In \S\ref{sect:dielectric} we show that the dispersion equations in ${\cal K}$ and in ${\cal K}'$ are proportional to each other. This allows us to adopt the convention that the wave modes are defined in the plasma rest frame, where there are only three wave modes (Paper~1). The wave modes in ${\cal K}'$ are Lorentz-transformed version of the corresponding modes in ${\cal K}$, and with our convention there are only three modes in ${\cal K}'$. However, as we show in \S\ref{sect:fourth}, each of these modes in ${\cal K}'$ may split into two branches; such splitting occurs when a positive-frequency, forward-propagating solution in ${\cal K}$ transforms into a negative-frequency or backward-propagating solution in ${\cal K}'$. An alternative convention would be to interpret each such split branch as two separate modes. The identification by \cite{BGI93} of four modes (rather than our three) in the pulsar frame may be interpreted as the longitudinal mode splitting into two. While we acknowledge that the number of modes is a matter of convention, in this case we argue that the specific forms of the dispersion relations derived by \cite{BGI93} (near $z'=1$) are based on an inappropriate approximation, effectively to the RPDF.

\section*{Acknowledgments} 
We thank Mike Wheatland and Axel Jessner for helpful comments on the manuscript, and referee V.~S.~Beskin and an anonymous referee for specific criticisms that led to significant revision of an earlier version of the manuscript.
The research reported in this paper was supported by the Australian Research Council through grant DP160102932.

\bibliographystyle{jpp}

\bibliography{Pulsar_radio_Refs}

\end{document}